\documentclass[%
 aip,
 amsmath,amssymb,
 reprint,%
]{revtex4-1}
\usepackage{graphicx}% Include figure files
\usepackage{dcolumn}% Align table columns on decimal point
\usepackage{bm}% bold math
\usepackage[utf8]{inputenc}
\usepackage[T1]{fontenc}
\usepackage{mathptmx}

\usepackage{hyperref} % produces nice hyperlinks in your document
\usepackage{braket}
\usepackage{amsthm}
\usepackage{tikz}
\usetikzlibrary{positioning}
\usetikzlibrary{angles,quotes}
\usetikzlibrary{arrows.meta,
                calc, chains,
                quotes,
                positioning,
                shapes.geometric}
\usepackage{caption}
\usepackage [english]{babel}
\usepackage [autostyle]{csquotes}
\usepackage{enumitem}
\usepackage{siunitx}
\usepackage{rotating}

\usepackage[T1,T2A]{fontenc}

\tikzstyle{Party} = [rectangle, minimum width=1.5cm, minimum height=1cm,text centered, draw=black, fill=gray!30]
\tikzstyle{ProtoPub} = [rectangle, rounded corners, minimum width=1.5cm, minimum height=1cm,text centered, draw=black]
\tikzstyle{ProtoPriv} = [rectangle, rounded corners, minimum width=1.5cm, minimum height=1cm,text centered, draw=black, fill=gray!10]
\tikzstyle{PubInfo} = [circle,text centered, draw=black]
\tikzstyle{PrivInfo} = [diamond, minimum width=1.5cm, minimum height=1cm,text centered, draw=black, fill=gray!10]

\MakeOuterQuote{"}

\theoremstyle{definition}
\newtheorem{defn}[equation]{Definition}
\theoremstyle{remark}

\theoremstyle{remark}

\renewcommand{\L}{\Lambda}
\newcommand{\R}{\mathbb{R}}
\newcommand{\Z}{\mathbb{Z}}
\newcommand{\N}{\mathbb{N}}
\newcommand{\B}{\mathbf{B}}
\newcommand{\F}{\mathbb{F}}
\newcommand{\E}{\mathbb{E}}
\renewcommand{\P}{\mathsf{P}}

\newcommand{\T}{\mathsf{T}}
\newcommand{\multi}{\mathsf{MQ}(\F^n,\F^m)}

\begin{document}

\preprint{AIP/123-QED}

\title[Signing Information in the Quantum Era]{Signing Information in the Quantum Era}
% Force line breaks with \\

\author{K. Longmate}
\author{E.M. Ball}%
 %\email{Second.Author@institution.edu.}
\affiliation{ 
Physics Department, Lancaster University, Lancaster LA1 4YB, United Kingdom%\\This line break forced with \textbackslash\textbackslash
}%

\author{E. Dable-Heath}
\affiliation{%
Electrical and Electronic Engineering, Imperial College London, South Kensington, London SW7 2BU, United Kingdom%\\This line break forced% with \\
}%

\author{R.J. Young}
\affiliation{ 
Physics Department, Lancaster University, Lancaster LA1 4YB, United Kingdom%\\This line break forced with \textbackslash\textbackslash
}%

\date{\today}% It is always \today, today,

\begin{abstract}
Signatures are primarily used as a mark of authenticity, to demonstrate that the sender of a message is who they claim to be. In the current digital age, signatures underpin trust in the vast majority of information that we exchange, particularly on public networks such as the internet. However, schemes for signing digital information which are based on assumptions of computational complexity are facing challenges from advances in mathematics, the capability of computers, and the advent of the quantum era. Here we present a review of digital signature schemes, looking at their origins and where they are under threat. Next, we introduce post-quantum digital schemes, which are being developed with the specific intent of mitigating against threats from quantum algorithms whilst still relying on digital processes and infrastructure. Finally, we review schemes for signing information carried on quantum channels, which promise provable security metrics. Signatures were invented as a practical means of authenticating communications and it is important that the practicality of novel signature schemes is considered carefully, which is kept as a common theme of interest throughout this review.
\end{abstract}

\maketitle
\tableofcontents

\section{Introduction}
Physical signatures are marks made to identify or authenticate the creator of a message or artifact. Their precise origins are lost to history, but they are associated with some of the earliest records of pictographic scripts, dating back at least 5 millennia\cite{normanIntro}. The information and telecommunications revolution in the second half of the 20th century would not have happened without a practical means to authenticate messages, which led to the invention of digital signature schemes.\\
\newline
Schemes for signing digital information are a direct, albeit stronger, analogue to physical signatures; they seek to ensure (i) authenticity of any claim regarding a message sender’s identity, (ii) that the message has not been altered by any parties since the signing, and (iii) that the sender cannot refute that it was indeed them who signed the message. In chapter 2 we review the origins of digital signatures, their applications and vulnerabilities associated with assumptions made in their foundation.\\
\newline
Post-quantum cryptography focuses on building classical algorithms whose security is resistant to known capabilities of quantum algorithms. Post-quantum signature schemes build upon early work in digital signatures. In chapter 3 we review progress made in this field and look closely at the resources required to implement these emerging algorithms.\\
\newline
It has been shown that algorithms using information encodes on quantum states can be used for secure communication protocols that are not dependent upon unproven assumptions, but instead are provably secure within the laws of physics itself. Chapter 4 discusses the application of quantum information and communications to signature schemes.
\section{Classical Signatures}
The pursuit of secure digital signature schemes was of great importance in 20th-century cryptographic research. Digital signatures are considered to be a cryptographic primitive with widespread application and use, with legal precedence in some jurisdictions. Like their physical counterparts, digital signatures are indeed used to authenticate the sending of a message, but their strength as a primitive protocol does not end here. Since their introduction in 1976 by Whitfield Diffie and Martin Hellman, \cite{DiffieHellman} further applications have been found in the building of secure distribution schemes, digitally processed financial transactions, cryptocurrencies\cite{signCurr} and more. It is known that a primitive analogue to digital signatures was developed decades before Diffie and Hellman made their public contributions, with the earliest known notion of authentication by some form of digital signature being a challenge-response mechanism used by the US Air Force to identify friendly aircraft, as far back as 1952\cite{diffie1988first}. Remarkably, even national identity and national government systems can be built with digital signatures at their core, as witnessed in Estonia's use of blockchain-style security for their Identity Card system, and \emph{eResidency} scheme offered to International visitors and investors \cite{estoniaRef}.

\subsection{Digital Signatures and Security Formalised}
In the following section we define digital signature schemes and their relevant security notions, as well as providing formal schematics of simple implementations of such schemes from the literature.

\begin{defn}[Digital Signature Scheme]
A \textbf{digital signature scheme} is a cryptographic protocol consisting of two distinct algorithms:
\begin{itemize}
    \item A \emph{signing} algorithm, in which the signing party (Alice), given a message (and typically a private key), produces a signature
    \item A \emph{verification} algorithm, in which, given the message and signature, the verifier (Bob) either accepts or rejects Alice's claim of authenticity of the message
\end{itemize}
\end{defn}
Digital signatures can fall into one of two categories, based on parties involved:
\begin{itemize}
    \item \textbf{True signatures}: Requiring only two parties, Alice (the signer) and Bob (the receiver), true signature schemes involve the transmission of information directly from Alice to Bob, typically in the form of a message-signature pair and most often using asymmetric key cryptosystems (public key cryptography)
    \item \textbf{Arbitrated signatures} Requiring a trusted third party, Charlie (the arbiter), this type of scheme involves two distinct rounds of communications: Alice's communication to Charlie, and Charlie's communication to Bob. In this setup, Charlie provides verification to Bob, and the landscape is opened up for the use of symmetric key cryptosystems (private key cryptography).
\end{itemize}
\noindent Digital signature schemes are typically preceded by some form of key generation (and distribution if necessary), allowing us to express all signature schemes in terms of the following three steps:
\begin{itemize}
    \item GEN: A key generation algorithm producing a private key (or set of private keys) and, if necessary, public keys.
    \item SIGN: Signature generated with a signing algorithm, and sent to Bob.
    \item VER: Bob receives the signature, and follows a verification algorithm before deciding whether or not to trust Alice's claim.
\end{itemize}
\noindent For any signature scheme to be considered secure and trustworthy for use, we require the scheme to provide the following under any and all conditions:
\begin{enumerate}
    \item \textbf{Authenticity}: The receiver, Bob, when accepting a signature from Alice is convinced that the author of the message was indeed Alice.
    \item \textbf{Integrity}: The receiver, Bob, can have faith that the message has not been altered since it left Alice.
    \item \textbf{Non-repudiation}: Once a genuine signed message has left Alice, she has no way to convince Bob that she was in-fact \textbf{not} the author.
\end{enumerate}
\noindent One more property often sought in signature schemes, but not strictly required for security, is for the signature to be able to be transferred. A signature scheme satisfying the above three conditions will convince Bob that Alice is indeed the author of the message, but transferability provides the ability for Bob to convince a third party, Charlie, that the message is indeed from Alice, without compromising the security of the system. \\
Attacks on signature schemes are known to typically fall into one of the four following categories: \label{AttCats}
\begin{enumerate}
    \item \textbf{Key-only attack}: An adversary, Eve, knows only the public key of Alice (the signer)
    \item \textbf{Known-signature attack}: Eve has access to Alice's public key, and message-signature pairs produced by Alice.
    \item \textbf{Chosen-message attack}: Eve may choose a list of messages $(m_1 , m_2 , ... , m_l)$, which she will ask Alice to sign.
    \item \textbf{Adaptively-chosen-message attack}: Similar to the above, except Eve has knowledge to adaptively choose messages based on the resulting message-signature pair of the previously signed message, allowing her to perform cryptanalysis with greater precision.
\end{enumerate}
And we may describe the level of success achieved by Eve, from greatest success to least success, as follows:
\begin{enumerate}
    \item \textbf{Secret key knowledge}: Eve discovers all of the secret information (typically Alice's secret key).
    \item \textbf{Universal forgery}: Eve is able to forge the signature of any message, but lacks the secret key itself.
    \item \textbf{Selective forgeries}: Eve can forge the signature for some messages of her choosing, but cannot do this arbitrarily.
    \item \textbf{Existential forgery}: Eve may forge the signature for \emph{at least one} message, but lacks the ability to choose this message from the set of all possible messages.
    \item \textbf{Failure}: Eve finds out nothing about the secret information, and fails to forge a convincing signature for \emph{any} message.
\end{enumerate}
Clearly, Eve achieving universal forgery or above would render the signature scheme completely invalid, as she could go on to convince Bob (and other parties) that Alice has signed any message (or, at least fail to be rejected with utmost confidence). When discussing full security for a signature scheme, it is typical to demand it not allow any form of success, i.e., not even existential forgery, under any computing assumptions (or none).\\
 
That existential forgery is considered not permissible may seem a somewhat "strong" requirement; we could easily suppose that, given Eve's inability to choose a message, we could simply require a very large message space and propose that a message containing "gibberish" would not be accepted by Bob. In the case of sending email communications, this may, at first, seem suitable. It seems reasonable for Bob to expect Alice's message to make sense in their chosen language, and given Eve has no control over the message contents, we might expect her to have difficulty randomly selecting a perfectly coherent message. However, given a scenario in which Alice is simply sending a number, related to an amount in currency she is requesting Bob send her, an existential forgery would carry great threat! Eve might not be able to choose a precise amount, but it would be hard for Bob to label a string of integers as nonsensical.

\subsection{Asymmetric Cryptography and Digital Signatures}
With research in digital signatures growing alongside research in public key cryptography\cite{diffie1988first}, the majority of well-known and well-studied signature schemes arise from public key cryptosystems. These typically rely upon certain mathematical assumptions about the hardness of problems (signature schemes based on symmetric encryption are generally reserved for arbitrated set-ups). An often-seen method of building a signature scheme is as follows: Find some public key cryptosystem based on one-way functions or trap-door functions, generate a signature using Alice's private key in the system, and allow any party to verify that Alice indeed sent the message using the publicly-shared encryption key. Well-known cryptosystems used in such a way include RSA\cite{RSAMain}, ElGamal\cite{elgamal1985public}, Rabin\cite{rabin1979digitalized}, and Fiat-Shamir\cite{fiat1986prove}. We remark that (in a simplified manner), the main property that distinguishes a one-way function from a trap-door function is the existence of \emph{trap-door knowledge}, some secret that allows the (usually) hard to invert function to become easily invertible.

\subsubsection{Modular Exponentiation and the RSA Cryptosystem}
A variety of trapdoor functions can be built based on performing exponentiation modulo $n$, depending on how we choose $n$. The simple act of squaring modulo $n$ where $n = pq$ for some prime $p,q$ forms a trap-door function, in which the trap-door knowledge is the prime factors ($p$ and $q$). We can build further trap-door functions with different exponents by carefully choosing the exponent. Again, working in modulo $n$ such that $n = pq$ for large primes $p,q$, if we choose some $e$ such that $e$ is coprime with $\phi(n) = (p-1)(q-1)$ (the Euler totient function of $n$) we find that for any given $x$, $$c = x^e \ (\text{mod }{n})$$ is a trapdoor function, where the trapdoor is once again the prime factors $p,q$. This forms the basis of the RSA cryptosystem.
Whilst the work of Diffie and Hellman in 1976 may have built the theoretical bench on which research could seek to implement digital signatures, it was a later paper by Rivest, Shamir and Adleman that first exemplified a proof-of-concept on top of this workbench. The well-celebrated RSA paper \cite{RSAMain} published in 1978 marked an early showcasing of asymmetric cryptosystems, well establishing the idea of public-key cryptography in a format that is in widespread use today. Relying on exponentiation under some $n = pq$ for large primes $p$ and $q$, the RSA cryptosystem can be used to send encrypted data securely (under assumptions), and the same methodology can be used to implement a signing algorithm securely. For the basis of an RSA-Implemented cryptosystem, private keys and public keys must be created for use in the trap-door function, all of which is formalised as follows:
\begin{itemize}
    \item \textbf{Trap-door function}: In the case of RSA, we take the trap-door function on some bit-string $m$ to be $$\text{RSA}_{\{E,D\}}(m) = m^{e,d} \ (\text{mod }{n})$$ Call $\text{RSA}_{E}$ the \emph{RSA encryption function} using key $e$, and $\text{RSA}_{D}$ the \emph{RSA decryption function} using key $d$, defined below. We require that $n = pq$ for some large primes $p$ and $q$, and choose $e$ such that for $\phi(n) = \phi(pq) = (p-1)(q-1)$, we have $\gcd(e,\phi(n)) = 1$, where gcd means greatest common divisor and $1 < e < \phi(n)$.
    \item \textbf{Public Key}: RSA takes as its encryption key the above chosen value, $e$. $e$ is part of the \emph{public} information in the cryptosystem, along with our chosen $n$ for modular arithmetic.
    \item \textbf{Decryption key}: One calculates the decryption key, $d$, by determining the multiplicative inverse of $e$ mod $\phi(n)$, i.e., determining $d \equiv e^{-1} (\text{mod } \phi(n))$. The decryption exponent, along with the prime factors $p,q$, of $n$, are kept secret. $d$ can be easily calculated when $p$ and $q$ are known.
\end{itemize}
We denote a message as $m$, with $C$ being the resulting ciphertext following encryption on $m$, and $S$ a signature (generated from a message $m$). The leading motivation for the RSA cryptosystem is its use for easy encryption of a message. Anyone with knowledge of the publicly shared information ($e$ and $n$) can easily encrypt a message $m$ by performing $C = m^e \mod n$. The intended recipient of the secret message, Alice, can easily recover $m = C^d \mod n$, and anyone lacking knowledge of $d$ who intercepts $C$ will struggle to find $m$ from just the public knowledge. Loosely, the "hardness" of recovering $m$ without $d$ relies upon the hardness of discovering $d$ without knowledge of $p$ and $q$. We then see that, really, the security of the RSA cryptosystem reduces down to the intractability of factoring $n$ into its prime factors $p,q$. This form of problem reduction is seen throughout digital signatures, and indeed all of cryptography.\\
 
Whilst the above demonstrates RSA's use as a tool to allow any party to transmit secret messages to a recipient Alice, it is easy to use the same tools to allow Alice to sign a message that can be verified by any party. Suppose Alice seeks to send a message, $m$, to some party Bob who wishes to verify that this message was not sent by some third party. This can be achieved by both parties carrying out the following:
\begin{enumerate}
    \item  Utilising her secret decryption key, Alice can now compute $ \text{RSA}_D(m) = m^d \mod n = S_m$. 
    \item Alice sends the message $m$ to Bob, along with the associated signature, $S_m$. 
    \item Bob simply calculates $\text{RSA}_e(S_x) = S_x ^e \mod n = (m^{d})^{e} \mod n = m^{d\cdot e} \mod n = m$
\end{enumerate}
From this, it is clear Bob can be convinced that only Alice (or someone with Alice's secret decryption key $d$) could have sent this message. As $e$ and $n$ are public knowledge, any other party may also be convinced of this, allowing transferability of the signature.\\
 
This (simplified) view of RSA demonstrates many fundamentals of digital signatures within a classical framework: The need for a cryptosystem whose security we have good reason to believe in, even if it is not provable (the assumption of intractability presents issues for provable security, see \ref{Sec_Close}), the ability to use this cryptosystem for signing (or at least modifying the cryptosystem for signing) and the need to ensure the three pillars of security for digital signatures: authenticity, integrity, and non-repudiation, whilst also hoping for the (at times less essential) property of transferability.

\begin{center}
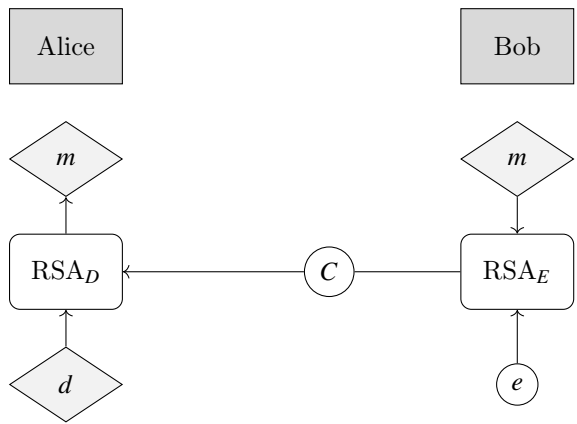

    \begin{tikzpicture}
    
    \node (Alice) [Party] {Alice};
    \node (AMessage) [PrivInfo, below of=Alice, yshift=-0.5cm] {$m$};
    \node (deRSA) [ProtoPub, below of=AMessage, yshift=-0.5cm] {$\textit{RSA}_D$};
    \node (deKey) [PrivInfo, below of=deRSA, yshift=-0.5cm] {$d$};
    \node (Code) [PubInfo, right of=deRSA, xshift =2.5cm] {$C$};
    
    \node (Bob) [Party, right of = Alice, xshift=5cm] {Bob};
    \node (BMessage) [PrivInfo, below of=Bob, yshift=-0.5cm] {$m$};
    \node (enRSA) [ProtoPub, below of=BMessage, yshift=-0.5cm] {$\textit{RSA}_E$};
    \node (enKey) [PubInfo, below of=enRSA, yshift=-0.5cm] {$e$};
    
    \draw [->] (BMessage) -- (enRSA); \draw [->] (enKey) -- (enRSA);
    \draw [->] (enRSA) -- (Code) -- (deRSA);
    \draw [->] (deKey) -- (deRSA);
    \draw [->] (deRSA) -- (AMessage);
    
    \end{tikzpicture}
    \captionof{figure}{Schematic demonstrating communications secured using the RSA protocol. Bob encrypts ($E$) a (private) message $m$, using the $RSA_E$ encryption function and the public key, $e$. The encrypted message, $C$, can now be sent publicly to Alice, who uses the $RSA_D$ decryption function and the private key $d$, to retrieve $m$. Circles represent information that can be presented publicly, whilst diamonds must remain private.}
\end{center}

\begin{center}
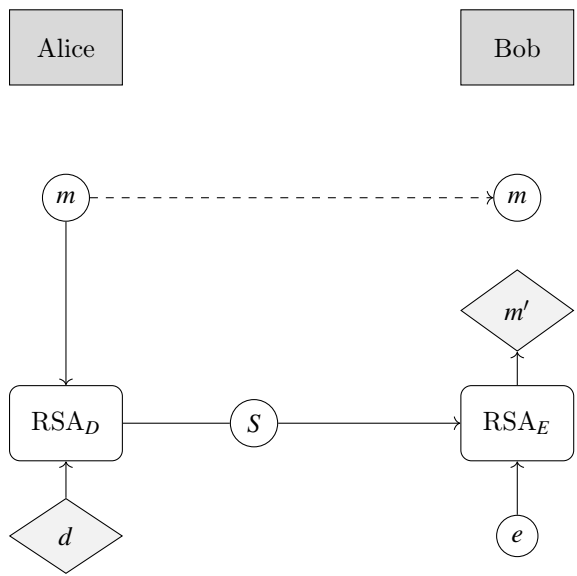

    \begin{tikzpicture}
    
    \node [Party] (Alice) {Alice};
    \node (AMessage) [PubInfo, below of=Alice, yshift=-1cm] {$m$};
    \node (deRSA) [ProtoPub, below of=AMessage, yshift=-2cm] {$\textit{RSA}_D$};
    \node (deKey) [PrivInfo, below of=deRSA, yshift=-0.5cm] {$d$};
    
    \node [Party, right of = Alice, xshift=5cm] (Bob) {Bob};
    \node (pubBMessage) [PubInfo, right of = AMessage, xshift=5cm] {$m$};
    \node (privBMessage) [PrivInfo, below of=pubBMessage, yshift=-0.5cm] {$m'$};
    \node (enRSA) [ProtoPub, below of=privBMessage, yshift=-0.5cm] {$\textit{RSA}_E$};
    \node (enKey) [PubInfo, below of=enRSA, yshift=-0.5cm] {$e$};
    \node (Signature) [PubInfo, left of=enRSA, xshift = -2.5cm] {$S$};
    
    \draw [->, dashed] (AMessage) -- (pubBMessage); \draw [->] (AMessage) -- (deRSA); \draw [->] (deKey) -- (deRSA);
    \draw [->] (deRSA) -- (Signature) -- (enRSA);
    \draw [->] (enKey) -- (enRSA);
    \draw [->] (enRSA) -- (privBMessage);
    
    \end{tikzpicture}
    \captionof{figure}{Schematic demonstrating message ($m$) signing using the RSA protocol. Alice computes a signature $S$ using the $RSA_D$ decryption function, her private key $d$ and a (public) message $m$. Both $m$ and $S$ can now be sent to Bob via a public channel. Bob can now compute $m'$ to be stored privately, using the $RSA_E$ encryption function, the retrieved signature $S$ and the public key $e$. If $m'$ matches $m$ closely (according to some pre-determined error-rate), the signature is accepted as valid. Otherwise, Alice is not accepted as the author of $m$. Circles represent information that can be presented publicly, whilst diamonds must remain private. Note the public and private variants of the message on Bob's end of communications. The private $m'$ is calculated from the signature, and its value is checked against the publicly sent $m$.}
\end{center}

\subsubsection{A note on symmetric digital signatures}
Whilst much of this section, along with the literature, focuses on asymmetric signature schemes whose roots lie in public key cryptosystems, this does not mean symmetric signature schemes arising from private key cryptosystems have no value in both research and application. Given the context (Alice signing a message, Bob verifying) it is easy to see why a private-key cryptosystem utilising the same (symmetric) key for both encryption and decryption is considered a weak arrangement for digital signatures: That both Alice and Bob use the same encryption key means either party can imitate the other. As long as Bob knows the encryption transformation used by Alice, he can always use the private key to generate a signature to deceive an unwitting third party, Charlie. The sought-after property of transferability is clearly lost. The potential applications for (secure) symmetric signatures is much smaller than that of asymmetric signatures. Both parties must trust each other to not be deceitful, which is far from practical for most settings (especially given the parties may have no knowledge of each other prior to the signing and verification). However, such systems are in use: for financial institutions they can be very beneficial as they are (often) less computationally taxing than their public key counterparts, and given a scenario where neither party has any reason to doubt the other's intentions (such as an ATM communicating with its parent financial institution), they can be used to great effect.

\subsection{Other Bases of Schemes}

\subsubsection{Modular Squaring}
The function $x \to x^2 \mod n$ for some $x \in \mathbb{Z}$ and $n = pq$, given $p,q$ are prime, forms a trap-door function in which the trap-door information, like in RSA, is knowledge of the prime factors $p,q$ of $n$. Rabin\cite{rabin1979digitalized} introduced a cryptosystem whose core is reliant on utilising squaring under modular arithmetic as a trap-door function. Use of Rabin's system, along with some hashing function, can produce a signature scheme with certain advantages over RSA. It is unknown whether or not breaking RSA is actually as difficult as factoring: the security reduction that reduces RSA to factoring remains unproven as the \emph{RSA assumption}. The potential that there may exist a security reduction from RSA to an easier problem than factoring is of concern. However, Rabin proved that breaking his cryptosystem is as difficult as the factoring problem. Thus, unlike RSA, cryptographers can find strength in its security as long as factoring remains intractible. 

\subsubsection{The Discrete Logarithm Problem}
Throughout this section, we have treated the RSA cryptosystem as a tool by which to lay out the general case for, and elucidate on general points within, digital signature schemes. Principally, we have opted for this approach as the RSA cryptosystem is a well-studied example of a cryptosystem built upon the notion of a trapdoor function. As we have previously covered, this constitutes a popular method of construction, allowing us to perform both encryption and signing. Yet not all public-key cryptosystems, or all cryptosystems used to deploy signature schemes, must be reliant upon trapdoor functions. The discrete logarithm problem is one such example of a one-way function with no trapdoor information that is successfully implemented in widely-used signature schemes. Working within finite fields, and letting $p$ be a prime and $g$ some primitive root in $\mathbb{Z}_p^*$, the function $$\text{dExp} : \mathbb{Z}_{p-1} \to \mathbb{Z}_p^*, \space x \mapsto g^x $$ is a one-way function with no trap-door knowledge. dExp (the discrete exponential function) is easy to compute under a finite field, but its inverse, dLog, is believed to be intractable (the discrete analogue of the logarithmic function is hard to compute, i.e., it is hard to find $x$ from $\text{dExp}(x) = g^x$) and there exists no "trap-door" information that makes this inverse easily computable. The assumption about the intractability of computing $x$ is known as the \emph{discrete logarithm assumption}, and can form the basis of a cryptosystem, most notably one devised by ElGamal \cite{elgamal1985public}, and has applications in signing of information.

\subsection{Security and Attacks across Asymmetric Signing}
%(Considered adding: - Zero-Knowledge signatures (Chaum) allowing choosing of verifiers. Jakobsson also good here: Designated verifiers for undeniable signatures)\\

\subsubsection{Signature Scheme Resistant to Adaptive Chosen Message Attacks}
Adaptively chosen-message attacks, as defined in \ref{AttCats}, utilise cryptanalysis of message-signature pairs (signed by Alice) as a powerful tool in a malicious party's (Eve's) pursuit of existential forgery against a given digital signature scheme. It is known that many well-studied cryptosystems are susceptible to such an attack, including RSA. In 1988 Goldwasser, Micali and Rivest (henceforth GMR) gave a thorough treatment of their signature scheme\cite{GMRMain}, whilst proving its security against adaptive chosen message attacks. Like many of the schemes preceding them, GMR's is reliant on trapdoors. However, GMR introduced the notion of \emph{claw-free permutation pairs}\cite{goldwasser1984paradoxical}, and claimed that signature schemes utilising trap-doors and claw-free permutations could produce an additional degree of security against adaptively chosen-message attacks when compared to the then-traditional method of `simple' trap-door schemes. Whilst the scheme itself isn't as simple and easily presentable as RSA, the basic notion behind the claw-free permutations is simple to see:
\begin{defn}
Given a set of numbers, $(x,y,z)$, we call them a \emph{claw} of two permutations $f_0$ and $f_1$ if $$f_0 (x) = f_1 (y) = z. $$ Further, we define a pair of permutations $f_0, f_1$ to be \emph{claw-free} if there exists no efficient algorithm for computing a claw given two permutations.
\end{defn}
GMR proved that the existence of such permutations implies the existence of a signature scheme $\epsilon$-secure against adaptively-chosen-message attacks, i.e., Eve achieves existential forgery with probability $< \epsilon$. Additionally, they presented a method of construction for practical claw-free permutations, utilising mathematical theory relevant to quadratic residues (an extensively studied tool in number theory, cryptosystems and cryptanalysis \cite{quadres}) in order to find piecewise functions $$f_0 (x) = g_0 (x) x^2 \ (\text{mod }{n})$$ and $$ f_1 (x) = g_1 (x) x^2 \ (\text{mod }{n})$$ where $g_0,g_1$ are piecewise constant functions. There exist functions in the form of $f_0, f_1$ that form claw-free permutations\cite{damgaard1987collision}. GMR show, via contradiction, that Eve's attempts of cryptanlaysis to achieve existential forgery can be reduced to finding a claw for the pair of permutations, and thus fail, even if the trap-door functions used independently remain vulnerable to adaptively-chosen-message attacks.\footnote{That this seems slightly paradoxical was certainly not lost on them, as seen by the title of their paper.}

\subsubsection{Hashing}
Typically when performing RSA with the RSA-encryption and decryption functions $RSA_{\{E,D\}}(m) = m^{\{d,e\}} \mod n$ with message $m$, encryption and decryption exponents $e,d$ respectively, and modulus $n$, we take $n$ to be some 1024-bit number. We bear in mind that, if sending a message in some text-based language, we are left with (at most) 128 ASCII characters. Assuming the language chosen is well-defined with a set of rules, we can assume most documents that need signing will be greater than this stringent limit. In order to allow the signing of messages and documents of arbitrary length, cryptographers typically turn to \emph{hash functions}.

\begin{defn}[Hash]
Simply, a \emph{Hash function}, $H$, is a function taking in as its input some data of arbitrary length, and outputting a \emph{hash digest} (or, simply, \emph{hash} or \emph{digest}) of a fixed length.
\end{defn}
For use in cryptography, we generally seek the following three properties from a hash function:
\begin{itemize}
    \item \textbf{Pre-image resistance}: Given a hash digest $h$, finding any message $m$ with $h = H(m)$ should be a difficult task. (We can consider the similarity between this property, and that of the one-way function.)
    \item \textbf{Collision resistance}: The essence behind collision resistance is that there should be a very low probability of  finding two messages outputting the same digest. Collision resistance is typically categorised into one of two groups: \emph{Weak collision resistance}, in which for given a message $m_1$, it should be difficult to find a message $m_2$ with $H(m_1) = H(m_2)$ when $m_1 \neq m_2$; and \emph{Strong collision resistance}, in which it should be difficult to find two messages $m_1 \neq m_2$ such that $H(m_1) = H(m_2)$.
\end{itemize}
Generally, it is favourable that these properties define a platform upon which a malicious adversary cannot modify the input data without changing the digest. Further, we desire a good distribution of digests, that is, given two n-bit-strings $m_1$ and $m_2$ with a small Hamming distance $\epsilon$, we seek very different outputs, i.e. a (relatively higher) Hamming distance between $H(m_1)$ and $H(m_2)$. Clearly, the overarching goal of creating a good hash function is that an adversary should find it very hard to determine the input of a hash, and cryptanalysis by attacks involving similar messages should be unable to find a weakness here.\\
 
Full security in the random oracle model can be achieved using a \emph{full domain hash function}, in which the image of the hash function is equal to the domain of the RSA function. However, most types of RSA widely used do not implement full-domain hash functions, instead opting for hash functions such as SHA, MD5, and RIPEMD \cite{penard2008secure, rivest1992md5, dobbertin1996ripemd}.

\subsubsection{Probabilistic Signatures}
In 1996, Bellare and Rogaway introduced the notion of the \emph{probabilistic signature scheme} (PSS)\cite{bellare1996exact}, in which the signature generated is dependent upon the message and a randomly chosen input. This results in a signature scheme whose output for a given message does not remain consistent over multiple implementations. Utilising a trap-door function (typically one well-used in non-probabilistic schemes, such as the RSA function), a hash function and some element of randomness (typically a pseudo-random bit generator), a signature scheme that is probabilistic in nature may be implemented. Such schemes can be used to sign messages of arbitrary length, and to ensure that the message $m$ is not recoverable from just the signature of $m$. RSA-PSS is a common probabilistic interpretation of the RSA signing scheme that forms part of the PKCS standards published by RSA laboratories\cite{PKCS1}.

\subsection{Cryptographic Standards and Modern Use}
We have already discussed how hashing may be used before signing a message (along with padding) to ensure all messages signed are of an appropriate size. However, the use of hashing in digital signatures extends beyond the "Hash and sign" idea used for signing protocols such as RSA. A protocol introduced by Fiat and Shamir \cite{fiat1986prove} has led to the creation of the \emph{Fiat-Shamir paradigm}. The Fiat-Shamir paradigm takes an interactive proof-of-knowledge protocol \cite{goldwasser1989knowledge} and replaces interactive steps with some random oracle, typically a publicly-known collision hash function. A thorough treatment of the paradigm can be found in Delfs' and Knebl's textbook on cryptography\cite{DelfsKneblBook}. In addition to their use in creating signature schemes that are secure against adaptively-chosen-message attacks, it has been shown by Damgard \cite{damgaard1987collision} that claw-free permutations can play a role in creating collision-resistant-hash-functions (this should not seem too surprising, as it is easily recognised that their definitions are similar: Collision - resistant hashing can almost be seen as a single-function analogue of claw-free permutations).\\
 
The work of ElGamal on cryptosystems making use of the one-way nature of the discrete logarithm forms the basis of the \emph{Digital Signature Algorithm}\cite{DSAnist}, a cryptographic standard popular since its proposal as a NIST submission for the \emph{Digital Signature Standard}, DSS.\\
  
Recent years have seen an increased interest in electronic voting, a concept heavily reliant on signature schemes. Electronic voting typically requires a cryptosystem that is both probabilistic and holds homomorphic properties. ElGamal is a good example of an applicable cryptosystem. Electronic voting has been used in a variety of countries, including the US(the 2000 Democratic Primary election in Arizona\cite{gibson2001elections} is often cited as a landmark event in internet voting); Scottish Parliament and local elections since 2007 (although the 2007 elections can be considered good proof as to why great care must go into researching the implementation of these systems before use \cite{kitcat2008observing}), Brazil\cite{fujiwara2015voting} (whose 2010 presidential election results were announced just 75 minutes after polls closed thanks to electronic voting), and India, with the state of Gujarat being the first Indian state to enable online voting in 2011 \cite{wolchok2010security}. In Europe, Estonia also utilise electronic voting \cite{estoniaRef}, with the idea of the Estonian digital ID-card, which provides a digital signature, being pivotal in how government and society are run in the Baltic country.\\
 
Another subfield of cryptographic research that has garnered increased interest in recent times is \emph{Elliptic Curve Cryptography}. Schemes based on the discrete logarithm problem (such as ElGamal/DSA) can be implemented similarly on the mathematical framework of elliptic curves\cite{koblitz1987elliptic} instead of finite fields. A key benefit of deploying a cryptosystem in such a way is the ability to perform computations at shorter binary lengths than traditionally used, without conceding security. This makes such schemes good candidates for when resources are limited, and Elliptic Curve DSA (ECDSA)\cite{johnson2001elliptic} is an example of such a scheme that forms a cryptographic standard, and is included in the Transport Layer Security (TLS) protocol.\cite{RFC4492} \\
 
In 1979, Ralph Merkle patented the concept of the \emph{hash tree}, commonly known as the \emph{Merkle tree} after him. Merkle trees can be paired with one-time signature schemes (within a symmetric cryptographic framework) to form a Merkle-Tree Based Signature scheme\cite{merkle1989certified}. Such schemes still remain only suitable for one-time use, although the work of Naor and Yung explores an extension of these types of schemes to complete multi-use signature schemes. It is believed that such signature schemes may be resistant to quantum-attacks, which are mentioned below and discussed further in section 2. 

\subsection{Looking forward for security}\label{Sec_Close}
As we have seen, the vast majority of widely implemented cryptographic algorithms (especially those that rise from public-key cryptosystems) rely upon unproven mathematical assumptions about the hardness of certain problems in order to provide us with security. This review is by no means expansive on the workings of different signature schemes under varied cryptosystems, and a reader seeking a thorough treatise of the field may turn to Simmons et al.\cite{SimmonsBook} (for an exploration of early public key cryptosystems and signatures) and Delfs-Knebl\cite{DelfsKneblBook} (for a treatise of modern cryptography, with extensive sections on signatures). With the increase in research in applications of quantum theory to modern technology, these previously held assumptions are left to fall apart in front of us. Since Deutsch's introduction of the \emph{Universal Quantum Computer}, \cite{deutsch1985quantum} research in utilising the power of quantum theory for computing has yielded many strong theoretical results, with early work including the development of algorithms for a quantum computer that can perform certain tasks faster than a classical computer is believed to be able to. Included in these is Shor's algorithm, \cite{ShorAlg} which can perform prime factorisation at a speed that would allow currently implemented cryptosystems to be broken. Whilst the practical implementation of such algorithms is yet to yield results strong enough to cause immediate worry, research is still looking forward to ensure security shall not be compromised as quantum computers grow more powerful.
\section{Post-Quantum Digital Signatures}

\subsection{Introduction: A Problem}
    As previously mentioned, advances in quantum computing have raised concerns for the field of classical cryptography. Here we give a brief overview of why, followed by a discussion of the responses from the cryptographic community.
    
    \subsubsection{Quantum Cryptanalysis of Classical Cryptography} 
        In an era where popular thinking was that problems based on factoring would be unbreakable, the introduction of Shor's algorithm \cite{ShorAlg} in 1994 caused uncertainty in the security of cryptosystems that were previously assumed to be secure. This review gives a brief overview of the techniques used. \\
         
        Factoring a composite number $ N $ can be reduced to the problem of finding the period of a function. This is done by picking a random number $ a<N $, checking that $ \gcd(a,N)=1 $ (if $ \gcd(a,N)\neq1 $ then we've found a factor of $ N $ and we're done), then looking for the period of the function:
        \begin{equation}
            f(x) = a^x\mod N.
        \end{equation}
        Up to this stage this can all be achieved classically. The quantum Fourier transform is used to find the period, resulting in Shor's algorithm being extremely efficient and appearing in the complexity class $ \mathsf{BQP} $ \cite{lin2014shor}. This is almost exponentially faster than the fastest known classical factoring algorithm, the general number field sieve \cite{pomerance2008tale}. \\
         
        This period solving algorithm can also be used to solve the discrete logarithm problem \cite{Shor99}, which also breaks the hardness assumption of this problem. From this, Shor's algorithm can be extended to a more general problem: the Hidden Subgroup Problem (HSP) \cite{jozsa2001quantum,grigni2001quantum,lomont2004hidden}. The HSP states that given a group $ G $, a finite set $ X $ and a function $ f:G\to X $ that \emph{hides} a subgroup $ H\leq G $, determine a generating set for $ H $ only given evaluations of $ f $. We say that a function $ f:G\to X $ hides $ H $ if, for all $ g_1, g_2\in G,~f(g_1) = f(g_2) $ if and only if $ g_1H=g_2H $. Within this framework, Shor's algorithm can be seen as solving the HSP for finite abelian groups. Other problems can similarly be generalised to this framework. For instance, if a quantum algorithm could solve the HSP for the symmetric group then one of the key hard problems for Lattice Cryptography (see section \ref{Lattice crypto}) - the shortest vector problem - would be broken \cite{DBLP:journals/corr/cs-DS-0304005}. \\
         
        Whilst Shor's algorithm has received the most attention for the problems it causes in cryptography, it is by far not the only quantum algorithm to attack current schemes. Grover's search algorithm \cite{grover1996fast,brassard1998quantum} can be used in certain schemes, and other factorisation algorithms such as the quantum elliptic-curve factorisation method \cite{bernstein2017post} have had some success. We point the reader in the direction of Bernstein et. al.\cite{bernstein2009introduction} and Jordan et. al.\cite{jordan2018quantum} for more complete surveys on quantum cryptanalysis of classical cryptography.
    
    \subsubsection{What's being done?}
        Whilst current estimates place the development of practical quantum computers capable of posing a security threat many years in the future (at time of writing the record for using Shor's algorithm to factor a `large number' into two constituent primes stands at $ 21 = 3\cdot 7~ $ \cite{Mart_n_L_pez_2012}, though much larger factorisations have been achieved in the adiabatic case \cite{li2017highfidelity}), it is pertinent to replace our current systems well in advance of that. In light of that, the National Institute for Standards and Technology (NIST) put out a call for submissions in 2016 \cite{NISTrnd2,alagic2019status} to attempt to set a new quantum-secure standard. This ongoing project aims to find new standards for both public key encryption and digital signatures.\\
        
        \paragraph{Author's Note} Since submitting this paper for review back in April NIST have announced that they have moved on to stage three of the standardisation process. This will be reflected in future drafts. \\
         
        The NIST evaluation criteria \cite{NISTcall} for these new schemes sets out both required security levels and computational cost. For the security levels, it is assumed that the attacker has access to signatures for no more than $ 2^{64} $ chosen messages using a classical oracle. The security levels are grouped into broad categories defined by easy-to-analyse reference primitives - in this case, the Secure-Hash-2 (SHA2) \cite{penard2008secure} and the Advanced Encryption Standard (AES) \cite{daemen1999aes}. The rationale behind the seemingly vague categories being that it is hard to predict advances in quantum computing and quantum algorithms, and so rather than using precise estimates of the number of `bits of security', a comparison will suffice. See table \ref{fig:NIST security levels} for the exact security levels. 
        
        \begin{table}[h!]
            \centering
            \begin{tabular}{c|c|c}
                Level & Reference Primitive & Security Equivalence \\
                \hline 
                \hline
                1 & AES 128 & Exhaustive key search \\
                2 & SHA 256 & Collision search \\
                3 & AES 192 & Exhaustive key search \\
                4 & SHA 384 & Collision search \\
                5 & AES 256 & Exhaustive key search
            \end{tabular}
            \caption{NIST security levels.}
            \label{fig:NIST security levels}
        \end{table}
        
        \noindent
        For quantum attacks, restrictions on circuit depth are given, motivated by the difficulty of running extremely long serial quantum computations. Proposed schemes are also judged on the size of the public keys and signatures they produce as well as the computational efficiency of the key generation.\\
         
        As of \today~the NIST project to set a new quantum-secure cryptographic standard is in the second round of submissions, with only nine proposals for digital signatures remaining. From these, it is clear that there are two front runners for the mathematical basis which will replace our current systems: Multivariate and Lattice cryptography. See figure \ref{Nist_Subs_tbl} for how these fall into the different categories. 
    n    
        \begin{table}[h!]
            \centering
            \begin{tabular}{c|c|c}
                 Multivariate & Lattice & Other \\
                 \hline
                 \hline 
                GeMSS \cite{GeMSSmain} & Dilithium \cite{Dilithiummain} & Picnic \cite{PICNICmain} \\
                LUOV \cite{LUOVmain} & FALCON \cite{FALCONmain} & SPHINCS$^+$ \cite{SPHINCSmain} \\
                MQDSS \cite{MQDSSmain} & qTESLA \cite{qTESLAmain} & \\
                Rainbow \cite{Rainbowmain} & & 
            \end{tabular}    
            \caption{NIST digital signature submissions by underlying mathematical structure type.}
            \label{Nist_Subs_tbl}
        \end{table}
         
        Here we will give a brief overview of both Multivariate and Lattice based cryptography, followed by a note on the other schemes.

\subsection{Multivariate Cryptography}
    Multivariate Cryptography was developed in the late 1980's with the work of Matsumoto and Imai \cite{imai1985algebraic}. Originally named $ C^* $ cryptography after the first protocol, the name Multivariate Cryptography was adopted when the work of Patarin broke and then generalised the $ C^* $ protocol \cite{patarin1995cryptanalysis}. After Shor's now infamous algorithm was developed it was realised that the structure of Multivariate Cryptography could be used as a direct response. For a more in depth treatment of the subject, the reader may consult Bernstein et. al.\cite{bernstein2009introduction} or Wolf\cite{wolf2006introduction}.\\
     
    All of the schemes are based on a hard problem which is relatively straightforward to understand (though several of the constructions extend the basic problem to more complex settings): the problem of solving multivariate quadratic equations over a finite field. That is, given a system of $ m $ polynomials in $ n $ variables:
    \begin{equation}\label{poly vector}
        \mathsf{P} = \begin{cases}
                        y_1 = p_1(x_1,\ldots, x_n) \\
                        y_2 = p_2(x_1, \ldots, x_n) \\
                        \vdots \\
                        y_m = p_m(x_1, \ldots, x_n)
                    \end{cases},
    \end{equation}
    and the vector $ \mathbf{y} = (y_1,\ldots, y_m)\in\F^m $, find a solution $ x\in\F $ which satisfies the equations above. Formally we say that for a finite field $ \F $ of size $ q:=|\F| $, an instance of an $ \mathsf{MQ}(\F^n,\F^m) $-problem is a system of polynomial equations of the form:
    \begin{equation}\label{multivariate quadratic polynomials}
        p_i(x_1,\ldots,x_n) = \sum_{1\leq j\leq k\leq n}\gamma_{ijk}x_j x_k + \sum_{j=1}^n \beta_{ij}x_j + \alpha_i,
    \end{equation}
    where $ 1\leq i\leq m $ and $ \gamma_{ijk},\beta_{ij},\alpha_{i}\in\F $. These are collected in the polynomial vector $ \P:=(p_1,\ldots,p_m) $. Here, $ \mathsf{MQ}(\F^n,\F^m) $ denotes a family of vectorial functions $ \P:\F^n\to\F^m $ of degree 2 over $ \F $:
    \begin{equation}
        \multi = \left\{\P = (p_1,\ldots,p_m) \mid \text{for} ~ p ~ \text{of the form \eqref{multivariate quadratic polynomials}}\right\}
    \end{equation}
    Whilst theoretically the polynomials could be of any degree, there is a trade-off between security and efficiency. Higher degrees naturally have larger parameter spaces, but too low a degree would be too easy to solve and therefore would be deemed too insecure. Quadratics are chosen as a compromise between the two.\\
     
    The final two pieces required for signatures are two affine maps, $ S\in\text{Aff}^{-1}(\F^n) $ and $ T\in\text{Aff}^{-1}(\F^m) $. Both of these can be represented in the usual way:
    \begin{align}
        S(x) &= M_S x + v_S \\
        T(x) &= M_T x + v_T
    \end{align}
    where $ M_S\in\F^{n\times n},~M_T\in\F^{m\times m} $ are invertible matrices and $ v_S\in\F^n,~v_T\in\F^m $ are vectors. \\
         
    For most multivariate signature schemes, the secret key is the triple $ (S^{-1},\P'^{-1},T^{-1}) $, where $ S $ and $ T $ are affine transforms and $ \P' $ is a polynomial vector (defined similarly to equation \eqref{poly vector}), known as the central equation. The choice of the shape of this equation is largely what distinguishes the different constructions in multivariate cryptography. The public key is then the following composition:
    \begin{equation}
        \P = S\circ\P'\circ T.
    \end{equation}
    To forge the signature, one would have to solve the following problem: \emph{for a given $ \P\in\mathsf{MQ}(\F^n,\F^m) $ and $ \mathbf{r}\in\F^m $ find, if any, $ \mathbf{s}\in\F^n $ such that $ \P(\mathbf{r})=\mathbf{s} $}. It was shown in Lewis et. al.\cite{lewis1983michael} that the decisional form of this problem is $ \mathsf{NP} $-hard, and it is believed to be intractable in the average case \cite{yasuda2015mq}.\\
     
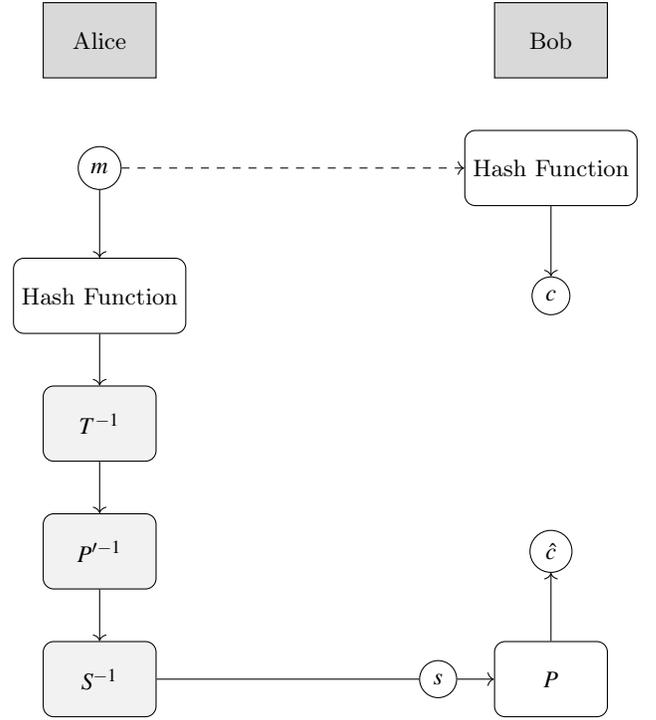
\begin{figure}
        \begin{center}
            \begin{tikzpicture}
            
            \node (Alice) [Party] {Alice};
            \node (AMessage) [PubInfo, below of=Alice, yshift=-0.7cm] {$ m $};
            \node (aHash) [ProtoPub, below of=AMessage, yshift=-0.7cm] {Hash Function};
            \node (privT) [ProtoPriv, below of=aHash, yshift=-0.7cm] {$ T^{-1} $};
            \node (privP) [ProtoPriv, below of=privT, yshift=-0.7cm] {$ P'^{-1} $};
            \node (privS) [ProtoPriv, below of=privP, yshift=-0.7cm] {$ S^{-1} $};
            
            \node (xSign) [PubInfo, right of=privS, xshift=3.5cm] {$ s $};
            
            \node (Bob) [Party, right of = Alice, xshift=5cm] {Bob};
            \node (bHash) [ProtoPub, below of=Bob, yshift=-0.7cm] {Hash Function};
            \node (hY) [PubInfo, below of=bHash, yshift=-0.7cm] {$ c $};
            \node (pubP) [ProtoPub, right of = privS, xshift=5cm] {$ P $};
            \node (yHat) [PubInfo, above of=pubP, yshift=0.7cm] {$ \hat{c} $};
            
            \draw [->, dashed] (AMessage) -- (bHash); \draw [->] (AMessage) --  (aHash);
            \draw [->] (aHash) -- (privT); \draw [->] (privT) -- (privP); \draw [->] (privP) -- (privS);
            
            \draw [->] (privS) -- (xSign) -- (pubP);
            
            \draw [->] (pubP) -- (yHat);
            \draw [->] (bHash) -- (hY);
            
            \end{tikzpicture}
        \end{center}
        \caption{A general schematic for Multivariate Quadratic signature schemes. Alice hashes the message $ m $ to some vector $ c\in\F^n $, which is transformed under the affine transform $ c'=T^{-1}(c) $. The central equation is then applied, $ s'=\mathsf{P}'(c') $, followed by a second affine transformation to create the signature $ s=S^{-1}(s') $. To check the signature Bob only has to recover $ \hat{c}=\mathsf{P}(s) [= S(\mathsf{P}'(T(s)))] $ and confirm that it matches the hash value $ H(m)=c $.}
        \label{fig:Multivariate Sig Scheme}
    \end{figure}

    We now formally outline the general scheme for signatures based on the Multivariate Quadratic problem:
    \begin{enumerate}[label = \roman*)]
        \item Alice generates a key pair $ (\mathsf{s}_k,\mathsf{p}_k) $, where $ \mathsf{s}_k = (S^{-1},\P'^{-1},T^{-1}) $ and $ \mathsf{p}_k=\P=S\circ\P'\circ T $, then distributes $ \mathsf{p}_k $.
        \item Alice then hashes the message, $ m $, to some $ c\in\F^n $ using a known hash function, then computes:
        \begin{equation*}
            s = \P^{-1}(c) = T^{-1}(\P'^{-1}(S^{-1}(c))),
        \end{equation*}
        sending the pair $ (m,s) $ to Bob.
        \item Bob then needs to check that $ \P(s) = c = H(m) $ for the known hash function $ H $.
    \end{enumerate}
    \noindent
    See figure \ref{fig:Multivariate Sig Scheme} for a diagram of how the signature schemes work. This forms the backbone of most multivariate schemes. The following subsections examine several of the adaptations of this framework employed in various NIST submissions.
    
    \subsubsection{Unbalanced Oil and Vinegar}
        The Oil and Vinegar scheme was first introduced by Patarin\cite{patarin1997oil}, but was broken by Kipnis and Shamir \cite{kipnis1999unbalanced} and generalised to the now common Unbalanced Oil and Vinegar (UOV) protocol. 
        
        \begin{defn}[Unbalanced Oil and Vinegar]
            Let $ \F $ be a finite field, $ o,v\in\N $ such that $ o+v=n $ and $ \alpha_i,\beta_{ij},\gamma_{ijk}\in\F $ for $ 1\leq i\leq v $ and $ 1\leq j\leq k \leq n $. Polynomials of the following form are central equations in the UOV-shape:
            \begin{equation}
                p_i(x_1,\ldots,x_n) := \sum_{j=1}^v\sum_{k=1}^n \gamma_{ijk}x_jx_k+\sum_{j=1}^n\beta_{ij}x_j + \alpha_i.
            \end{equation}
            The first $ x_1,\ldots,x_v $ terms are known as the vinegar terms and the second register of $ o = n-v $ terms are called the oil terms. If $ o\neq v $ it is called \emph{unbalanced}. 
        \end{defn}
        
        In these equations, the vinegar terms are combined quadratically with themselves, and then combined quadratically with the oil terms, whereas the oil terms are never mixed with themselves. For a secure construction, the required discrepancy between the number of oil and vinegar terms is $ v\geq 2o $. Unbalanced Oil and Vinegar has become one of the most common constructions for Multivariate Cryptography, and it has itself become a way of varying other constructions by putting them in an UOV-shape. Two of the NIST submissions are directly based on UOV: LUOV and Rainbow. One of the major problems with UOV is the length of the signatures and the key sizes, and both of these submissions get around this by introducing additional structure on top of the UOV shape. For a comparison of signature and key size as well as other efficiency markers, see section \ref{Crypto Performance}.

    \subsubsection{Hidden Field Equations}
        The Hidden Field Equations (HFE) protocol \cite{patarin1995cryptanalysis} is a generalisation of one of the original multivariate systems, the Matsumoto-Imai scheme \cite{imai1985algebraic}. Similar to Oil and Vinegar, the underlying scheme was broken before the underlying trapdoor was generalised. However, unlike UOV, this scheme uses more than one field: the ground field $ \F $ and it's $ n^{th} $-degree field extension $ \E $, that is $ \E:=\F[t]/f(t) $ where $ f(t) $ is an irreducible polynomial over $ \F $ of degree $ n $. 
        
        \begin{defn}[Hidden Field Equations (HFE)]
            Let $ \F $ be a finite field with $ q:=|\F| $ elements, $ \E $ its $ n^{th} $-degree extension field and $ \phi:\E\to\F^n $ the canonical, coordinate-wise bijection between the extension field and the vector space. Let $ P(X) $ be a univariate polynomial over $ \E $ with:
            \begin{equation}
                P'(X) := \underset{q^i+q^j\leq d}{\sum_{0\leq i,j\leq d}}C_{ij}X^{q^i+q^j} + \underset{q^k\leq d}{\sum_{0\leq k\leq d}} B_kX^{q^k} + A,
            \end{equation}
            where
            \begin{align*}
                C_{ij}X^{q^i+q^j} &\quad \text{for} ~ C_{ij}\in\E ~ \text{are quadratic terms}, \\
                B_kX^k &\quad \text{for} ~ B_k\in\E ~ \text{are linear terms, and} \\
                A &\quad \text{for} ~ A\in\E ~ \text{is constant,}
            \end{align*}
            for $ i,j\in\N $ and a degree $ d\in\N $. We say that central equations of the form $ \P':=\phi\circ P'\circ\phi^{-1} $ are in HFE shape.
        \end{defn}
        
        The GeMSS submission to the NIST proceedings uses Hidden Field Equations, although it adapts the form using `minus and vinegar modifiers' \cite{wolf2002hidden}. This has allowed the design to become more flexible in its choice of security parameters whilst improving efficiency. 
        
    \subsubsection{5-Pass SSH}
        Unlike the other schemes mentioned, the MQDSS NIST submission uses the Fiat-Shamir paradigm to develop a digital signature from a secure identification scheme, namely the 5-pass Sakumoto, Shirai and Hiwatari (SSH) identification scheme \cite{sakumoto2011public} (5-pass here refers to 5 passes of information between the prover and verifier. They also included a 3-pass scheme, but the MQDSS team found it to be inefficient as a signature scheme \cite{hulsing20165}.) . Here we give a brief overview of the identification scheme followed by its adaptation to a signature scheme.\\
         
        The SSH scheme does not rely on structure built on top of the $ \mathsf{MQ} $ problem, which allows them to avoid said structure from being taking advantage of for cryptanalysis, as in the cases of Patarin\cite{patarin1996hidden}, Thomae\cite{thomae2013security} and Faugere et. al.\cite{faugere2008cryptanalysis}. Instead it is based on an alternative form of the central equation known as the \emph{polar form}.
        
        \begin{defn}[Polar form]
            Let $ \P\in\multi $ and $ \mathbf{r},\mathbf{s}\in\F^n $, then the \emph{polar form} of $ \P $ is:
            \begin{equation}
                \mathsf{T}(\mathbf{r},\mathbf{s}) := \P(\mathbf{r}+\mathbf{s}) - \P(\mathbf{r}) - \P(\mathbf{s})
            \end{equation}
        \end{defn}
        
        Sakumoto et al. use the polar form to represent the secret as $ \mathbf{s}:= \mathbf{t}_0 + \mathbf{t}_1 $ so that the public part can be written as $ \mathbf{r} := \P(\mathbf{t}_0) + \P(\mathbf{t}_1) + \T(\mathbf{t}_0,\mathbf{t}_1) $. They then further represent $ \mathbf{t}_0 $ and $ \P(\mathbf{t}_0) $ as $ \alpha\mathbf{t_0}=\mathbf{u_0}+\mathbf{u_1} $ and $ \alpha\P(\mathbf{t}_0=\mathbf{v}_0+\mathbf{v}_1 $ where $ \alpha $ is randomly selected from $ \F $. Taking advantage of the linearity of the polar form $ \mathbf{r} $ can be represented as:
        \begin{equation}
            \alpha \mathbf{r} = (\mathbf{v}_1+\alpha\P(\mathbf{t}_1)+\T(\mathbf{u}_1,\mathbf{t}_1)) + (\mathbf{v}_0+\T(\mathbf{u}_0,\mathbf{t}_1),
        \end{equation}
        thus obfuscating the secret. They further hide the secret by using a commitment scheme prior to the intial pass. It was proven in Sakumoto et. al.\cite{sakumoto2011public}, that when the commitment scheme used was statistically hiding the whole scheme is statistically zero-knowledge.\\
         
        Using the Fiat-Shamir heuristic, the 5 interactive passes of the identity scheme are replaced by non-interactive random-oracle access, specifically by a series of hash functions that replicate the challenges. They go on to prove the security of this scheme in the random oracle model in H\"{u}lsing et. al.\cite{hulsing20165}
    
    \subsubsection{Attacks}
        The cryptanalysis of multivariate schemes comes in two forms:
        \paragraph{Structural} These focus on taking advantage of the specific structural faults in the design on different protocols. Included amongst this are attacks on a form of Multivariate cryptography called MINRANK \cite{faugere2008cryptanalysis} and the hidden field equations \cite{dubois2007cryptanalysis}.
        
        \paragraph{General} Attacks that directly try and break the underlying hardness assumption of solving multivariate equations. These include the use of techniques such as utilising Gr\"{o}bner bases to make the solving of the multivariate systems easier. For a good overview of the area see Billet and Ding\cite{billet2009overview}.

\subsection{Lattice Cryptography}\label{Lattice crypto}
    Lattice cryptography was first introduced by the work of Ajtai \cite{ajtai1996generating} who suggested that it would be possible to base the security of cryptographic systems on the hardness of well-studied lattice problems. The familiarity of these problems made them an attractive candidate for PQC. This led to the development of the first lattice-based public-key encryption scheme - NTRU \cite{hoffstein1998ntru}. However, this was shown to be insecure and it would take the work of Regev to establish the first scheme whose security was proven under worst-case hardness assumptions \cite{regev2009lattices}. For an overview of the field of lattice cryptography, we direct the reader to Peikert\cite{peikert2016decade}.\\
         
    There is a whole suite of lattice problems on which cryptographic schemes are based. Here - following some basic definitions - we will introduce the key ideas that form the foundation of contemporary Lattice Cryptography.
        
    \begin{defn}
        A \emph{lattice} $ \L\subset\R^n $ is a discrete additive subgroup of $ \R^n $. That is $ \mathbf{0}\in\L $, if $ \mathbf{x},\mathbf{y}\in\L $ then $-\mathbf{x},\mathbf{x}+\mathbf{y}\in\L $, and any $ \mathbf{x} \in\L $ has a neighbourhood of $ \R^n $ which has no other lattice points.
    \end{defn}
    
    We note here that lattices can be more generally defined as a discrete additive subgroup of some general vector space $ V $, but are most commonly restricted to $ \R^n $. Any non-trivial lattice is countably infinite, however each lattice can be finitely generated by all the integer combinations of some set of vectors in $ \R^n $, $ \B=\{\mathbf{b}_1,\ldots,\mathbf{b}_k\} $ for some $ k\leq n $. Typically $ k=n $, in which case we call $ \L $ a full rank lattice. We call $ \B $ the basis for a lattice $ \L $ and express it as a matrix of row vectors. Often we describe the lattice as a linear sum of integer multiplied basis vectors, writing:
        
    \begin{equation}\label{lattice bases}
        \L(\B) = \left.\left\{\sum_{i=1}^k z_i\mathbf{b}_i \right| z_i\in\Z\right\}.
    \end{equation}
    \noindent
    A basis $ \B $ isn't unique as any two bases $ \B_1,~\B_2 $ for a lattice $ \L $ are related by a unimodular matrix $ \mathbf{U} $ such that $ \B_1 = \mathbf{U}\B_2 $. Another crucial lattice definition is the dual lattice:
    
    \begin{defn}
        Given a lattice $ \L $ in $ V $ where $ V $ is endowed with some inner product $ \langle\cdot,\cdot\rangle $, the dual lattice $ \L^{\perp} $ is defined as $ \L^{\perp}=\{\mathbf{v}\in V \mid \langle\L,\mathbf{v}\rangle\subset\Z\} $.
    \end{defn}
    
    Here we give a quick note about additional structure that can be imbued in lattices. A common technique is to construct lattices embedding algebraic structures, such as rings and modules. It is beyond the scope of this paper to go into detail on how one constructs these, so we point the readers in the direction of Lubashevksy et. al.\cite{lyubashevsky2010ideal} and Grover\cite{CharlieThesis} for a more complete understanding of the structure. The justification for these will become clear once we introduce lattice hard problems.\\
         
    Another important concept is that of Discrete Lattice Gaussians:
        
    \begin{defn}[Discrete Lattice Gaussian]
        Given a basis $ \B $ for a lattice $ \L(\B) $, mean $ \mu\in\R^n $ and standard deviation $ \sigma > 0 $, the discrete Gaussian over a lattice is defined as,
        \begin{equation}
            D_{\L, \sigma, \mu}(\mathbf{x}) := \frac{\rho_{\sigma,\mu}(\B\mathbf{x})}{\rho_{\sigma,\mu}(\L)}, \quad \mathbf{x}\in\Z^n,
        \end{equation}
        where $ \rho_{\sigma,\mu} (\mathbf{y}):=\exp\left(-\frac{1}{2\sigma^2}\|\mathbf{y}-\mu\|\right) $ and $ \rho_{\sigma, \mu} (\L) = \sum_{\mathbf{x}\in\Z^n} \rho_{\sigma,\mu}(\B\mathbf{x}) $.
    \end{defn} 
    
    Discrete lattice Gaussian sampling is one of the core features of Lattice Cryptography, being employed in some manner in most schemes. However, this form of sampling comes with a whole host of issues. For one, it is computationally hard to sample directly from such distributions, leading to algorithms that sample from statistically close distributions \cite{klein2000finding}. Unfortunately, these approximate distributions aren't necessarily spherical Gaussians and therefore have the potential to leak information about the secret \cite{howe2016practical}. Other inherent problems include finding the upper and lower bounds on the choice of variance - too low a variance also leaks information, but too high a variance will produce signatures that are insecure (see later sections). See Prest\cite{prest2015gaussian} for a more comprehensive discussion on discrete Gaussian sampling.
        
    \subsubsection{Fundamental Hard Problems}
        We now move on to the hard lattice problems. First, we give a few of the fundamental hard problems, which form a foundation for the hard problems that contemporary lattice cryptography is built on.
    
        \begin{defn}[Shortest Vector Problem $ (\mathsf{SVP}) $]
            Define $ \lambda_1(\L) $ to be the length shortest non-zero vector in $ \L $. \emph{The Shortest Vector Problem} $ (\mathsf{SVP}) $ is: given a basis $ \mathbf{B} $ of a lattice $ \L $, compute some $ \mathbf{v}\in\L $ such that $ \|\mathbf{v}\| = \lambda_1(\L) $.
        \end{defn}
        
        Here, $ \lambda_m(\L) $ are the succesive minima of the lattice, where each vector $ \|v_i\| =  \lambda_i \leq \lambda_j $ for $ i < j $, with $ \lambda_1(\L) $ being the shortest vector in the lattice. The norm function $ \|\cdot\| $ is left intentionally unspecified, though it is typically the Euclidean norm. This problem is regarded as hard in both a classical and quantum setting, but it falters when applied to cryptographic schemes with some probabilistic element. It is more common to use the approximate analogue to the $ \mathsf{SVP} $, which is as follows:
        
        \begin{defn}[Approximate Shortest Vector Problem ($\mathsf{SVP}_{\gamma} $)] 
            Given a basis $ \mathbf{B} $ of a lattice $ \L(\B) $, find a non-zero vector $ \mathbf{v}\in\L $ such that $ \|\mathbf{v}\|\leq\gamma(n)\cdot\lambda_1(\Lambda) $.
        \end{defn} 
        
        We also make mention of \emph{bounded distance decoding} ($\mathsf{BDD}$) which asks the user to find the closest lattice vector to a prescribed target point $ \mathbf{t}\in\L $ which is promised to be `rather close' to the lattice.
        
        \begin{defn}[Bounded Distance Decoding ($ \mathsf{BDD}_{\gamma} $)] 
            Given a basis $ \B $ of a full-rank lattice $ \L(\B) $ and a target vector $ \mathbf{t}\in\R^n $ with a guarantee that $ \emph{dist}(\mathbf{t},\L) < d = \lambda_1(\L)/(2\gamma(n)) $, find unique lattice vector $ \mathbf{v} \in \L $ such that $ \|\mathbf{t}-\mathbf{v}\|<d $.
        \end{defn}
        
        The above problems have varying degrees of provable hardness. The exact version of the shortest vector problem is known to be NP-hard under randomised reductions \cite{ajtai1998shortest}, however, the implementation of the hard problems such as bounded distance decoding relies on polynomial encoding so it is in the complexity class $\mathsf{NP}\cap\mathsf{co-NP} $ \cite{aharonov2005lattice}. Currently, there are no known quantum algorithms that solve any of the above problems in polynomial time, but there have been various attempts, see section \ref{Lattice Cryptanalysis} for further detail.
    
    \subsubsection{Foundations of Contemporary Lattice Crypto}
        Whilst the previous hard problems are fundamental to lattices, they are not easily implementable in lattice schemes. Here we introduce the two problems which form the foundation for contemporary lattice cryptography: \emph{the short integer solution} $ (\mathsf{SIS}) $ \cite{ajtai1996generating} and \emph{learning with errors} $ (\mathsf{LWE}) $ \cite{regev2009lattices}.
        
        \begin{defn}[Short Integer Solution ($ \mathsf{SIS}_{n,q\beta,m} $)]
            Given $ m $ uniformally random vectors $ \mathbf{a}_i\in\Z_q^n $ forming the columns of a matrix $\mathbf{A}\in\Z_q^{n\times m} $, find a nonzero integer vector $ \mathbf{z}\in\Z^m $ of norm $ \|\mathbf{z}\|\leq\beta $ such that:
            \begin{equation}
                f_\mathbf{A}(\mathbf{z}) := \mathbf{A}\mathbf{z} = \sum_i \mathbf{a}_i\cdot z_i = \mathbf{0} \in\Z_q^n.
            \end{equation}
        \end{defn}
        
        $ \mathsf{LWE} $ is an average-case problem introduced by Regev, often referred to as the `encryption enabling' analogue of the SIS problem.
        
        \begin{defn}[$ \mathsf{LWE} $ distribution]
            For a vector $ \mathbf{z}\in\Z^n_q $ called the \emph{secret}, the $ \mathsf{LWE} $ distribution $ A_{\mathbf{z},\chi} $ over $\Z_q^n\times\Z_q $ is sampled by choosing $ \mathbf{a}\in\Z_q^n $ uniformly at random, choosing $ e\longleftarrow\chi $, and outputting $ (\mathbf{a}, b=\langle\mathbf{z},\mathbf{a}\rangle+e\mod q) $.
        \end{defn}
        
        The problem comes in two distinct forms: \emph{decision} and \emph{search}. Decision requires distinguishing between $ \mathsf{LWE} $ samples and uniformly random ones, whereas search requires finding a secret given $ \mathsf{LWE} $ samples.
        
        \begin{defn}[Decision $ \mathsf{LWE}_{n,q,\chi,m} $]
            Given $ m $ independent samples $ (\mathbf{a}_i,b_i)\in\Z_q^n\times\Z_q $ where every sample is distributed according to either:
            \begin{enumerate}[label = \roman*)]
                \item $ A_{\mathbf{z},\chi} $ for a uniformly random $ \mathbf{z}\in\Z_q^n $ (fixed for all samples),
                \item The uniform distribution,
            \end{enumerate}
            distinguish which is the case.
        \end{defn}
        
        \begin{defn}[Search $ \mathsf{LWE}_{n,q,\chi,m} $]
            Given $ m $ independent samples $ (\mathbf{a}_i, b_i)\in\Z_q^n\times\Z_q $ drawn from $ A_{\mathbf{z},\chi} $ for a uniformly random $ \mathbf{z}\in\Z_q^n $, find $ \mathbf{z} $.
        \end{defn}
        
        The learning with errors problem has been shown to be at least as hard as \emph{quantumly} solving $ \mathsf{SVP} $ on \emph{arbitrary} $ n $-dimensional lattices. The following security reduction can be shown following the proof from Regev\cite{regev2009lattices}:
        
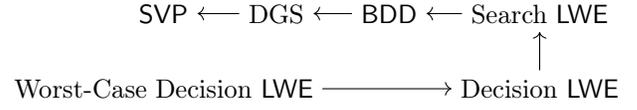
        \begin{center}
            \begin{tikzpicture}
                \node   (SVP)                                         {$ \mathsf{SVP} $};
                \node   (DGS)   [right of=SVP, xshift=5mm]                         {DGS};
                \node   (BDD)   [right of=DGS, xshift=5mm]              {$\mathsf{BDD}$};
                \node   (SLWE)  [right of=BDD, xshift=10mm]      {Search $\mathsf{LWE}$};
                \node   (DLWE)  [below of=SLWE]                 {Decision $\mathsf{LWE}$};
                \node   (WLWE)  [below of =SVP]    {Worst-Case Decision $\mathsf{LWE}$};
                
                \draw[->] (DGS) -- (SVP);
                \draw[->] (BDD) -- (DGS);
                \draw[->] (SLWE) -- (BDD);
                \draw[->] (DLWE) -- (SLWE);
                \draw[->] (WLWE) -- (DLWE);
            \end{tikzpicture}    
        \end{center}
        \noindent
        Here DGS stands for discrete Gaussian sampling. We note that the reduction between discrete Gaussian sampling and the $ \mathsf{BDD} $ problem is a quantum step.\\
         
        As previously mentioned, it is possible to construct lattices from specific algebraic structures such as rings or modules \cite{albrecht2017large}, when combined with the above problems it is known as structured-$ \mathsf{LWE} $ or structured-$ \mathsf{SIS} $. This is largely done for efficiency reasons as the parameter space needed to implement systems based on these structures is greatly reduced \cite{lyubashevsky2010ideal}. The choice of which structure is worth using has some nuances, however. For example, ring-$ \mathsf{LWE} $ is generally considered to be more efficient than module-$ \mathsf{LWE} $, however the efficiency comes at a cost in flexibility and security. Increasing the security of a scheme requires increasing the dimension of the lattice, which in the ring case is often chosen such that $ N=n $ where $ n=2^k $ for some integer $ k $. Thus, going up a security level requires going from dimension 512 to 1024 for instance, whereas a more optimal scheme may lie inbetween these. Module-$ \mathsf{LWE} $ has dimension parametrised by an integer $ d $ such that $ N = dn $, again for $ n $ of the form $ 2^k $. Setting $ d=3 $ and $ n=256 $ allows for a total dimension of 768, which may be preferable for the targeted level of security. Even further structure can be imbued which may yet give greater flexibility: middle-product-$ \mathsf{LWE} $ \cite{rocsca2017middle} and cyclic-$ \mathsf{LWE} $ \cite{grovernon}. The security of all of these schemes based on algebraic structure has been questioned however, as the reduction to standard LWE is not fully understood \cite{peikert2016not,langlois2015worst,roux2014lattice}.\\
         
        All of the lattice based NIST submissions for signature schemes use some kind of structure: qTESLA and FALCON are based on rings (FALCON, however, uses a distinction between binary and ternary forms to capture the intermediate security levels) whereas Dilithium is based on module-$ \mathsf{LWE} $.

    \subsubsection{The GPV Framework}
        Introduced in the seminal paper of Gentry et. al. \cite{gentry2008trapdoors}, the Gentry-Peikert-Vaikuntanathan (GPV) framework gives an overarching structure for taking advantage of `natural' trapdoors in lattices to obtain signatures. It is built on a signature scheme first introduced in Goldreich-Goldwasser-Halevi (GGH) \cite{goldreich1997public} and NTRUsign \cite{hoffstein2003ntrusign} schemes:
        \begin{itemize}
            \item The public key is a full rank matrix $ \mathbf{A}\in\Z_q^{n\times m} $ generating a lattice $ \L $. The private key is a matrix $ \B\in\Z_q^{m\times m} $ generating $ \L_q^{\perp} $, the dual lattice of $ \L\mod q $.
            \item Given a message $ m $, a signature is a short value $ \mathbf{s}\in\Z_q^m $ such that $ \mathbf{s}\mathbf{A}^T = H(m) = \mathbf{c} $ where $ H:\{0,1\}^*\to\Z_q^n $ is a known hash function. Given $ \mathbf{A} $, verifying $ \mathbf{s} $ as a valid signature is straightforward: check $ \mathbf{s} $ is short and that $ \mathbf{s}\mathbf{A}^T=\mathbf{c} $.
            \item Computing a signature requires more care however:
            \begin{enumerate}[label = \roman*)]
                \item Compute an arbitrary preimage $ \mathbf{c}_0\in\Z_q^m $ such that $ \mathbf{c}_0\mathbf{A}^T = \mathbf{c} $. $ \mathbf{c}_0 $ is not required to be short so it can be computed with relative ease.
                \item Use $ \B $ to compute a vector $ \mathbf{v}\in\L_q^{\perp} $ close to $ \mathbf{c}_0 $. Then $ \mathbf{s}=\mathbf{c}_0-\mathbf{v} $ is a valid signature:
                \begin{equation}
                    \mathbf{s}\mathbf{A}^T = \mathbf{c}_0\mathbf{A}^T-\mathbf{v}\mathbf{A}^T = \mathbf{c}-\mathbf{0} = \mathbf{c}.
                \end{equation}
                If $ \mathbf{c}_0 $ and $ \mathbf{v} $ are close enough then $ \mathbf{s} $ will be short, fulfilling the second requirement of a valid signature.
            \end{enumerate}
        \end{itemize}
        
       \begin{figure}[h!]
        \centering

        \begin{tikzpicture}
            \node (Alice) [Party] {Alice};
            \node (message) [PubInfo, below of=Alice, yshift=-0.75cm] {$m$};
            \node (aHash) [ProtoPub, below of=message, yshift=-0.75cm] {Hash Function};
            \node (outHash) [PubInfo, below of=aHash, yshift=-0.5cm] {$c$};
            \node (leftVec) [ProtoPriv, below of=outHash, yshift=-0.625cm, xshift=-1cm] {$VS$};
            \node (privB) [PrivInfo, left of=outHash, xshift=-0.75cm] {$B$};
            \node (rightVec) [ProtoPriv, below of=outHash, yshift=-0.625cm, xshift=1cm] {$PC$};
            \node (aliceA) [PubInfo, right of =outHash, xshift=0.75cm] {$A$};
            \node (vector) [PrivInfo, below of=leftVec, yshift=-0.5cm] {$v$};
            \node (preimage) [PrivInfo, below of=rightVec, yshift=-0.5cm] {$c_0$};
            \node (signature) [PubInfo, below of=outHash, yshift=-3.25cm] {$s$};
    
            % Right Hand-side
    
            \node (Bob) [Party, right of=Alice, xshift=3.75cm] {Bob};
            \node (bHash) [ProtoPub, below of=Bob, yshift=-0.75cm] {Hash Function};
            \node (bobHash) [PubInfo, below of=bHash, yshift=-0.75cm] {$c$};
            \node (ver) [ProtoPub, right of=signature, xshift=3.75cm] {Ver};
            \node (cHat) [PrivInfo, above of=ver, yshift=1.75cm] {$\hat{c}$};
            \node (bobA) [PubInfo, below of=ver, yshift=-0.5cm] {$A$};
    
            % Arrow drawing
    
            \draw [->, dashed] (message) -- (bHash); \draw [->] (message) -- (aHash);
            \draw [->] (aHash) -- (outHash); \draw [->] (outHash) -- (leftVec); \draw [->] (outHash) -- (rightVec);
            \draw [->] (privB) -- (leftVec); \draw [->] (aliceA) -- (rightVec);
            \draw [->] (leftVec) -- (vector); \draw [->] (rightVec) -- (preimage);
            \draw [->] (preimage) -- (signature); \draw [->] (vector) -- (signature);
    
            \draw [->] (signature) -- (ver); draw [->] (bobA) -- (ver); \draw [->] (ver) -- (cHat); \draw [->] (bHash) -- (bobHash);
            \draw[->] (bobA) -- (ver);
    
        \end{tikzpicture}
        \caption{A schematic of the GPV signature framework. Alice hashes the message to a lattice vector $ \mathbf{c}=H(m) $. Following this, she creates the key pair $ (pk,sk)=(\mathbf{A}\in\Z^{n\times m}_q,\B\in\Z^{m\times m}_q) $ and sends the public key to Bob. Using elementary techniques, she computes the preimage vector (PC) $ \mathbf{c}_0\mathbf{A}=\mathbf{c} $, then using the secret key she samples the vector (VS) $ \mathbf{v}\in\L^{\perp} $ such that it is very close to $ \mathbf{c}_0 $. The signature is $ \mathbf{s}=\mathbf{c}_0-\mathbf{v} $. Bob has to check the signature is small enough and, using the public key, that $ \mathbf{s}\mathbf{A}^T[=\mathbf{c_0}\mathbf{A}^T-\mathbf{v}\mathbf{A}^T]=H(m) $. Here $ \mathbf{v}\mathbf{A}^T=\mathbf{0} $ since $ \mathbf{A} $ generates the lattice and $ \mathbf{v} $ belongs to the dual lattice.}
        \label{fig:GPV framework schematic}
        \end{figure}
        \noindent
        The GGH and NTRUsign schemes, however, proved insecure as the method for computing vector $ \mathbf{v}\in\L^{\perp} $ leaked information about secret basis $ \B $ and have since been proven to be insecure by cryptanalysis \cite{nguyen1999cryptanalysis,gentry2002cryptanalysis,ducas2012learning}. The GPV framework differs in that instead of the deterministic algorithm - Babai's roundoff algorithm - it uses Klein's algorithm \cite{klein2000finding}, a randomised variant of the nearest plane algorithm also developed by Babai\cite{babai1986lovasz}. Whereas both deterministic algorithms would leak information about the geometry of the lattice, Klein's avoids this by sampling from a spherical Gaussian over the shifted lattice $ \mathbf{c}_0+\L $.\\
          
        Klein's algorithm was the first of a family of lattice algorithms known as trapdoor samplers. The GPV framework has become a generic framework into which a choice of trapdoor sampler and lattice structure can be inserted. It has also been proven to be secure under the assumptions of $ \mathsf{SIS} $ under the random oracle model \cite{gentry2008trapdoors,katsumata2018tighter}.\\
         
        A prudent example of an instantiation of the GPV framework is the NIST submission FALCON \cite{fouque2018falcon}. This uses a trapdoor sampler known as the Fast-Fourier-Sampler - developed by Prest and Ducas \cite{ducas2016fast} - over NTRU lattices that take advantage of the ring structure.
    
    \subsubsection{Bai-Galbraith Signatures}
        The Bai-Galbraith signature scheme \cite{bai2014improved} is an adaptation of an earlier work by Lyubashevesky \cite{lyubashevsky2009fiat} in which he develops a lattice-based signature scheme using a paradigm which he calls `Fiat-Shamir-with-aborts'. Informally, it follows the chain of reductions:
        \begin{center}
            \begin{tikzpicture}
                \node   (HP)                                          {Hard Problem};
                \node   (CRHF)  [right of=HP, xshift=10mm]                    {CRHF};
                \node   (OTS)   [right of=CRHF, xshift=15mm]    {One-time signature};
                \node   (IDS)   [below of=OTS]                           {ID Scheme};
                \node   (Sig)   [below of=HP]                            {Signature};
                
                \draw[->]   (CRHF) -- (HP);
                \draw[->]   (OTS) -- (CRHF);
                \draw[->]   (IDS) -- (OTS);
                \draw[->]   (Sig) -- (IDS);
            \end{tikzpicture}
        \end{center}
        Here CRHF stands for collision resistant hash function. The main idea is that, from a lattice based CRHF, one can create a one-time signature following Lyubashevsky and Micciancio\cite{lyubashevsky2008asymptotically}, however this leaks information about the secret key. This would not be a problem for a one-time signature as the information becomes defunct after the usage. Constructing the ID scheme requires the repeated use of this, which is where the aborting technique comes in. In response to the challenge from the verifier, the prover can decide that sending the usual response would leak information and instead abort the protocol and restart it. The end result is a secure, if somewhat inefficient, ID scheme. Having to restart the protocol every time there is information leaked is done away with when adapting this to a signature scheme using Fiat-Shamir, however. The lack of interaction means that the prover can simply rerun the protocol until they find a signature which does not leak information. \\
         
        As an $ \mathsf{LWE} $ instantiation, Lyubashevesky's scheme has public key $ (\mathbf{A},\mathbf{b} = \mathbf{A}\mathbf{z}+\mathbf{e}\mod q) $, where the components are picked as described above. The verifier picks small-normed vectors $ \mathbf{y}_1 $, $ \mathbf{y_2} $ and computes $ \mathbf{v} = \mathbf{A}\mathbf{y}_1+\mathbf{y}_2 $. With the message $ m $ they compute the hash $ \mathbf{c}:=H(\mathbf{v},m) $ and the following two vectors: $ \mathbf{s}_1=\mathbf{y}_1+\mathbf{z}\mathbf{c} $ and $ \mathbf{s}_1=\mathbf{y}_2+\mathbf{e}\mathbf{c} $. The signature is then $ s =  (\mathbf{s}_1,\mathbf{s}_2,\mathbf{c}) $. Here, to ensure that neither $ \mathbf{s}_1 $ or $ \mathbf{s}_2 $ leak information about the protocol, rejection sampling is employed. Developed across \cite{ducas2013lattice,guneysu2012practical,lyubashevsky2012lattice} this technique allows the vectors to be picked from a distribution independent of the secret. Verification requires checking that $ \|\mathbf{s}_1\| $ and $ \|\mathbf{s}_2\| $ are small enough, and that $ H(\mathbf{A}\mathbf{s}_1 + \mathbf{s}_2 - \mathbf{b}\mathbf{c}\mod q, m) = \mathbf{c} $. This can be thought of as a proof of knowledge of $ (\mathbf{z},\mathbf{e}) $.\\
         
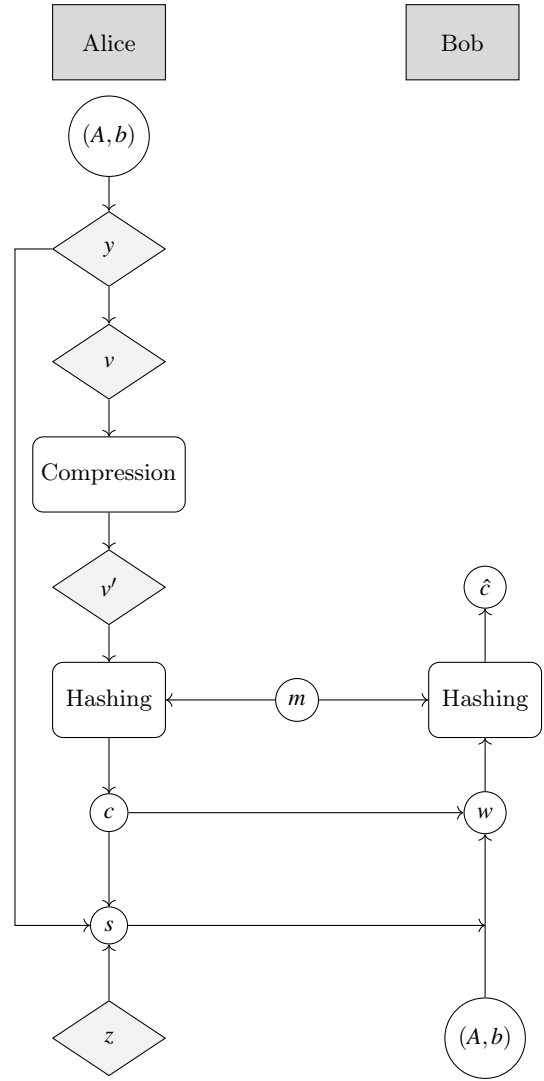
\begin{figure}[h!]
            \centering
            \begin{tikzpicture}
            
                \node (Alice) [Party] {Alice};
                \node (pubkey) [PubInfo, below of=Alice, yshift=-0.25cm] {$ (A,b) $};
                \node (y) [PrivInfo, below of=pubkey, yshift=-0.5cm] {$ y $};
                \node (vector) [PrivInfo, below of=y, yshift=-0.5cm] {$ v $};
                \node (comp) [ProtoPub, below of=vector, yshift=-0.5cm] {Compression};
                \node (altv)  [PrivInfo, below of=comp, yshift=-0.5cm] {$ v' $};
                \node (aHash) [ProtoPub, below of=altv, yshift=-0.5cm] {Hashing};
                \node (message) [PubInfo, right of=aHash, xshift=1.5cm] {$m$};
                \node (cAlice) [PubInfo, below of=aHash, yshift=-0.5cm] {$ c $};
                \node (s)   [PubInfo, below of=cAlice, yshift=-0.5cm]  {$ s $};
                \node (secret)   [PrivInfo, below of=s, yshift=-0.5cm]  {$ z $};
                
                \node (Bob) [Party, right of=Alice, xshift=3.7cm] {Bob};
                \node (cBob) [PubInfo, right of=altv, xshift=4cm] {$\hat{c}$};
                \node (bHash) [ProtoPub, right of=aHash, xshift=4cm] {Hashing};
                \node (bCalc) [PubInfo, right of=cAlice, xshift=4cm]  {$ w $};
                \node (pubA) [PubInfo, right of=secret, xshift=4cm] {$(A,b) $};
                
                \draw [->] (pubkey) -- (y); \draw [->] (y) -- (vector);
                \draw [->] (vector) -- (comp);
                \draw [->] (comp) -- (altv);
                \draw [->] (altv) -- (aHash);
                \draw [->] (aHash) -- (cAlice);
                \draw [->] (cAlice) -- (s);
                \draw [->] (secret) -- (s);
                \draw [->] (message) -- (aHash);
                \draw [->] (cAlice) -- (bCalc);
                
                \draw [->] (s) -- ++ (5,0);
                \draw [->] (y) -- ++ (-1.25,0) -- ++ (0,-9) -- (s);
                
                \draw [->] (message) -- (bHash); \draw [->] (bHash) -- (cBob);
                \draw [->] (pubA) -- (bCalc); \draw [->] (bCalc) -- (bHash);
            
            \end{tikzpicture}
            \caption{A schematic for Bai-Galbraith Signatures. Alice first generates the key pair $ (pk,sk) = ((\mathbf{A},\mathbf{b}=\mathbf{A}\mathbf{z}+\mathbf{e}\mod q),\mathbf{z}) $, sending the public key to Bob. After picking a short vector $ \mathbf{y} $, she finds the finds the lattice vector $ \mathbf{v}=\mathbf{A}\mathbf{y} $ and removes the most significant bits to compute $ \mathbf{v}'=comp(\mathbf{v}) $. She then hashes this with the message to create the vector $ \mathbf{c} $ from hash value $ c=H(\mathbf{v}',m) $. The signature is $ \mathbf{s}=\mathbf{y}+\mathsf{z}\mathbf{c} $. Bob computes the vector $ \mathbf{w}=\mathbf{A}\mathbf{s}-\mathbf{b}\mathbf{c} $ and checks that the hash value of the most significant bits of $ \mathbf{w} $ and the message $ m $ is consistent with $ c $.}
            \label{fig:Bai-Galbraith}
        \end{figure}

        Bai and Galbraith adapted this such that it instead becomes a proof of knowledge of only $ \mathbf{z} $ using a variation of the verification equation and compression techniques. Once the public key has been created, only one vector is required, $ \mathbf{y} $, from which $ \mathbf{v}=\mathbf{A}\mathbf{y}\mod q $. The least significant bits of $ \mathbf{v} $ are then thrown away and the remainder is hashed with the message $ m $ to get a hash value $ c $. From this value, the vector $ \mathbf{c} $ is created and the signature is $ (\mathbf{s}=\mathbf{y}+\mathbf{z}\mathbf{c},c) $, with rejection sampling again being used to check that the distribution of $ \mathbf{z} $ is independent of the secret. Computing $ \mathbf{w}=\mathbf{A}\mathbf{s}-\mathbf{b}\mathbf{c}\equiv\mathbf{A}\mathbf{y}-\mathbf{e}\mathbf{c}\mod q $ allows for verification by checking that the hash value of the most significant bits of $ \mathbf{w} $ with $ m $ is equal to $ c $. See figure \ref{fig:Bai-Galbraith} for a schematic for the Bai-Galbraith signature scheme.
        
        \noindent
        Two of the current NIST submissions are based on this signature scheme: Dilithium and qTESLA. qTESLA uses the ring-$ \mathsf{LWE} $ variant whereas Dilithium is based on module-$ \mathsf{LWE} $.
    
    \subsubsection{Attacks}\label{Lattice Cryptanalysis}
        Similar to the multivariate case, there is a breadth of literature on specific attacks, both quantum and classical, and suffice to say we will not be going into them in too much depth. Here we give a brief summary of some of the avenues that have been attempted and references for readers to investigate. Similar to attacks on Multivariate Cryptographic schemes, these can largely be broken down into three categories: attacks on the underlying hard problems, attacks on specific schemes and side-channel attacks (see section \ref{side channel attacks} for details of side-channel attacks). \\ 
    
        \paragraph{General structure attacks} The security of lattice cryptography can be largely reduced down to the shortest vector problem, so the general motivation for these attacks is to solve said problem. Schemes of this kind tend to come in two forms: algorithmic or sampling. Largely, the question to be answered is the following: given a random basis for a lattice, can one find the shortest vector? Or at least a small enough vector, satisfying the approximate $ \mathsf{SVP} $? On the algorithmic side, there are lattice reduction algorithms such as Schnorr\cite{schnorr2009progress} or Lyu et. al.\cite{lyu2018lattice} which attempt to find almost-orthogonal bases from a highly non orthogonal bases. In a similar vein, quantum speed ups of vector enumeration have been proposed by Aono et. al.\cite{aono2018quantum} There are also search approaches such as the sieving algorithms \cite{laarhoven2015sieving} and their quantum counterparts \cite{laarhoven2015finding}. Newer approaches, devised using adiabatic quantum computing, posing $ \mathsf{SVP} $ as an energy minimisation problem as in Joseph et. al.\cite{PhysRevResearch.2.013361, joseph2020quantum} have also been developed. Similar to the use of sampling to generate small signatures, if a truly efficient discrete Gaussian sampler was developed it could pose a major problem \cite{ajtai2002sampling,aggarwal2015solving}. Simply setting the centre of the distribution as the zero vector would allow the shortest vector to be picked with a high probability. There have also been suggestions that certain quantum algorithms could be used to pick vectors more efficiently from these distributions. \\
         
        As mentioned previously, the $ \mathsf{SVP} $ can be shown to be equivalent to solving the Hidden Subgroup Problem for symmetric groups, which has also been the focus of much quantum cryptanalytic research \cite{suo2020quantum}.\\
        
        \paragraph{Specific attacks} For details on how individual schemes are taking known attacks into account, we refer the reader to the design documents: Dilithium \cite{dilithiumdesign}, FALCON \cite{FALCONdesign}, qTESLA \cite{qtesladesign}.
        
\subsection{Symmetric Primitive Based Submissions}
    Here we give a brief overview of the remaing two NIST submissions, Picnic and SPHINCS$ ^+ $, both of which are based on symmetric primitives.
    
    \subsubsection{Picnic}
        Unlike the previously mentioned schemes, Picnic only requires the hardness provided by symmetric primitives such as hash functions and block ciphers \cite{chase2017post}. It is a general scheme for the adaptation of a three-move proof-of-knowledge scheme (known as $ \Sigma $-protocols) to signature using a transformation from Unruh \cite{unruh2012quantum}. It is claimed by Unruh \cite{unruh2015non} that the Fiat-Shamir paradigm for transforming proof-of-knowledge schemes into signatures is impractical to prove secure in the quantum random oracle model and so Unruh provides an alternative. The Picnic protocol provides two signature schemes: one via Unruh and another using Fiat-Shamir. \\
         
        Picnic is built upon a $ \Sigma $-protocol called ZKB++ \cite{chase2017post}, which itself is built on an earlier scheme called ZKBOO \cite{giacomelli2016zkboo}. For the sake of brevity, the details of Picnic and the underlying schemes are omitted and instead we will explain the underpinning framework by first explicitly defining $ \Sigma $-protocols followed by Unruh's transformation.
        
        \begin{defn}[$ \Sigma $-protocol]
            A three-move proof-of-knowledge protocol between a prover (Alice) and a verifier (Bob) is known as a \emph{$ \Sigma $-protocol}. Alice wants to prove she knows $ x $ such that $ f(x)=y $, where $ y $ is commonly known, for some relation $ f $.
            \begin{enumerate}
                \item Alice commits herself to randomness by picking $ r $, which she sends to Bob.
                \item Bob replies with a random challenge $ c $.
                \item Alice responds to the challenge with a newly computed $ t $.
                \item Bob accepts that Alice has proven the knowledge if $ \phi(y,r,c,t)=1 $ for some efficiently computable and agreed upon $ \phi $.
            \end{enumerate}
        \end{defn}
        
        Unruh's transform takes a given $ \Sigma $-protocol with a challenge space $ C $, an integer $ N $, message $ m $ and a random permutation $ G $ and requires the following:
        \begin{enumerate}
            \item Alice runs the first phase of the $ \Sigma $-protocol $ N $ times to produce $ r_1,\ldots,r_N $.
            \item For each $ i\in\{1,\ldots,N\} $, and for each $ j\in C $, she computes the responses $ t_{ij} $ for $ r_i $ and challenge $ j $. She then computes $ g_{ij}=G(t_{ij}) $.
            \item Using a known hash function, she computes $ H\left(x,r_1,\ldots,r_N,g_{11},\ldots,g_{N|C|}\right) $ to obtain indices $ J_1,\ldots,J_N $.
            \item The signature she outputs is then $ s=\left(r_1,\ldots,r_N,T_{1J_1},\ldots,T_{NJ_N},g_{11},\ldots,g_{N|C|}\right) $.
        \end{enumerate}
        Bob then verifies the hash, verifies that the given $ t_{iJ_i} $ values match the corresponding $ g_{iJ_i} $ values, and that the $ t_{iJ_i} $ values are valid responses with respect to the $ r_i $ values. \\
         
        Whilst Picnic is based on the $ \Sigma $-protocol ZKB++, there is some choice in the use of the symmetric primitives used in the construction. The choice that has been implemented in the block cipher family LowMC \cite{albrecht2015ciphers}, which is based on a substitution permutation network. Performance wise, Picnic is relatively slow and employs large signature sizes, but makes up for this with provable quantum security.
    
    \subsubsection{SPHINCS$ ^+ $}
        SPHINCS$ ^+ $ \cite{bernstein2019sphincs+} is based on an earlier protocol called SPHINCS \cite{bernstein2015sphincs}. This protocol is a hash-based, stateless signature scheme which had the goal of having the practical elements of other hash-based schemes and adding extra security by removing the stateful nature. A stateful algorithm depends in some way on a quantity called the \emph{state}, which is initialised in some way. This is often a counter, though not necessarily, and stateful schemes can lead to many insecurities as they need to keep track of all produced signatures. \\
         
        SPHINCS expands the idea of using a Merkle tree \cite{merkle1989certified} to extend a one-time signature into a many-time signature scheme by creating a hypertree. In this tree of trees, leaves of the initial Merkle tree become the root one-time signatures for further trees, which themselves cascade into further trees. The size of the overall tree becomes a compromise between security and efficiency in the original scheme: which leaves are picked to become the next tree are chosen randomly, and so a smaller tree has a chance to repeat the leaf choice. In order to abate this, a few-times signature is used at the bottom of the tree. This randomness is what makes SPHINCS stateless. Whereas Merkle's original design iterates over the signing keys, SPHINCS builds on the theoretical work of Goldreich\cite{goldreich1986two}, in which the keys are picked randomly. The size of the hypertree allows the assumption that the new key has not been used before. \\
         
        Improving on the previous design, SPHINCS$ ^+ $ uses a more secure few-time signature at the bottom of the tree, known as Forest of Random Subsets (FORS), an improvement on a previous signature called HORST \cite{bernstein2015sphincs}. A better selection algorithm for choosing the leaves of the tree is also included. It also introduces the idea of \emph{tweakable hash functions}.
        \begin{defn}[Tweakable hash functions.]
            Let $ \alpha, n\in\N,\mathcal{P} $ be the public parameters space and $ \mathcal{T} $ be the tweak space. A \emph{tweakable hash function} is an efficient function
            \begin{equation}
                TH:\mathcal{P}\times\mathcal{T}\times\{0,1\}^\alpha\to\{0,1\}^n, \quad MD\leftarrow TH(P,T,m),
            \end{equation}
            mapping an $ \alpha $-bit message $ m $ to an $ n $-bit hash value message digest (MD) using a function key called public parameter $ P\in\mathcal{P} $ and a tweak $ T\in\mathcal{T} $.    
        \end{defn}
        
        This allows the hash functions to be generalised to the whole hypertree as they can adapt to changes in the chosen leaves of the sub trees. In brief, the SPHINCS$ ^+ $ signature scheme can be summarised as follows:
        \begin{enumerate}
            \item Alice generates $ p,q\in\{0,1\}^n $: $ p $ is a seed for the root of the top tree in the hypertree and $ q $ is a public seed. The pair $ (p,q) $ form the public key. The secret key is the pair $ t,u\in\{0,1\}^n $, respectively seeds for the few-time signature FORS and the chosen one-time signature for the protocol, WOTS$ ^+ $ \cite{bernstein2019sphincs+}.
            \item To sign the message, Alice generates the hypertree and the signature is the following collection: a FORS signature on a message digest, a WOTS$ ^+ $ signature on the FORS public key, a series of authentication paths and WOTS$ ^+ $ signatures to authenticate the the WOTS$ ^+ $ public key.
            \item To verify this, Bob iteratively reconstructs the public keys and root nodes until the top of the SPHINCS$ ^+ $ hypertree is reached.
        \end{enumerate}
        The SPHINCS$ ^+ $ protocol was not designed with the same kind of performance in mind as either the lattice or multivariate schemes. Generally, stateless signatures have much larger key and signature sizes, as well as slower performances. They mainly target applications which have low latency requirements but very strong security requirements, such as offline code signing. In Bernstein et. al.\cite{bernstein2019sphincs+} the reader will find an analysis of the security of this scheme in both a classical and quantum setting that shows it to be very strong in both regards.
    
    \subsubsection{Attacks}
        Here we include references for cryptanalysis efforts of the above schemes. For Picnic the recent attacks include a multi-attack on the scheme and it's underlying zero-knowledge protocols \cite{cryptoeprint:2018:1212}, and an attack on the block cipher used in implementation, LowMC \cite{cryptoeprint:2018:859}. Both of these - as well as some side-channel attack analysis - are addressed in the design document \cite{PICNICdesign}.\\
         
        Currently, main attack against the SPHINCS framework is that of Castelnovi, Martinelli and Prest \cite{castelnovi2018grafting,genet2018practical}. This is a type of side-channel attack known as a differential fault attack. In the design document \cite{SPHINCSdesign}, general protection against this kind of attack - as well as other known general attacks - is addressed.

\subsection{Side-Channel Attacks}\label{side channel attacks}
    It would be remiss of this paper to give an overview of the state of contemporary cryptographic signatures - especially with regards to new standards - without a note on side-channel attacks. Side-channel attacks are cryptanalytic attacks that focus on finding flaws in the implementation of protocols rather than the design. Examples of this include timing attacks - where an adversary can gleam information about a secret from a protocol by taking advantage of a subroutine running in non-constant time- and energy attacks - a similar process but instead requires examining the energy use of the protocols. This has led to some of the bigger breaks of security systems that have been employed \cite[p.~116]{katz2014introduction}. Unfortunately, in the case of Post-Quantum Cryptography, it is often a form of attack which has not been considered in as much depth as is potentially necessary and many of the submissions are missing a large scale analysis of how they could be affected. Similar to the attacks previously mentioned, however, it is beyond the scope of this paper to go into a great amount of detail so instead we provide a list of references for invested readers. \\
         
    For a good overview of side-channel attacks in general, we refer the readers to Fan et. al.\cite{fan2010state} or Lo'ai et. al.\cite{lo2016towards} Beyond this, we direct the readers towards specific analysis of side channel attacks for certain NIST submissions: \\
     
    There has been some work on general fault attacks on Multivariate public key cryptosystems \cite{hashimoto2013general}. Side-channel attacks on both LUOV and Rainbow can be seen in the general attacks on Unbalanced Oil and Vinegar schemes \cite{yi2018side,kramer2019fault,ding2019new}. \\
     
    Dilithium was attacked using a side-channel assisted existential forgery attack \cite{cryptoeprint:2018:821}. This was responded to with countermeasures suggested in the implementation that mask the protocol \cite{cryptoeprint:2019:394}. FALCON has recently been attacked using a protocol known as BEARZ \cite{mccarthy2019bearz}. The attacking party also suggested countermeasures to prevent this fault attack and timing attacks on FALCON. Similarly, qTESLA has had implementation countermeasures suggested \cite{cryptoeprint:2019:606}. \\
     
    For each submission, we also direct the readers to the respective design documents \cite{GeMSSmain,dilithiumdesign,PICNICdesign,LUOVmain,FALCONdesign,SPHINCSdesign,MQDSSmain,qtesladesign,Rainbowmain}. Unfortunately the level of detail on each is not to an equal standard, with some severely lacking side-channel attack analysis.

\subsection{Performance}\label{Crypto Performance}
    When considering the performance of these different protocols, there are various angles to analyse them. At a top level view one could compare a range of properties such as the key size, length of signatures, verification times and signature creation times. NIST make the point that these algorithms will be employed in a multitude of applications, each with different requirements. For example, if the applications can cache public keys, or refrain from transmitting them frequently then the size of the public key is not as important. Similarly, in terms of the computational efficiency, a server with high traffic spending a significant portion of its resources verifying client signatures will be more sensitive to slower key operations. The call for proposals \cite{NISTcall} even suggests that it may be necessary to standardise more than one algorithm to meet the differing needs. \\
     
    Whilst the computational efficiency relies on the specific architecture used, the key and signature sizes can be compared theoretically. A comparison of the NIST schemes for these architectures can be found in table \ref{fig:postq Key and sig comp table}. Many of the submissions include data on several variants of their respective schemes but we have only included a cut down list here. The variants - as well as the original data - can be found in the submissions themselves\cite{MQDSSmain,Dilithiummain,PICNICmain,LUOVmain,FALCONmain,SPHINCSmain,MQDSSmain,qTESLAmain,Rainbowmain}. \\
    
    \begin{table}[h!]
        \centering
        \begin{tabular}{c||r|r|c}
            Submission & {PK Size} & {Signature Size} & Security level \\
            \hline
            \hline
            GeMSS 128 & 352188 & 32 & 1 \\
            GeMSS 192 & 1237964 & 51 & 3 \\
            GeMSS 256 & 3040700 & 72 & 5 \\
            \hline
            LUOV-8-58-237 & 12100 & 311 & 2 \\
            LUOV-8-82-323 & 34100 & 421 & 4 \\
            LUOV-8-107-371 & 75500 & 494 & 5 \\
            \hline
            MQDSS-31-48 & 20 & 6492 & 1/2 \\
            MQDSS-31-64 & 28 & 13680 & 3/4 \\
            \hline 
            Rainbow Ia & 148500 & 32 & 1 \\
            Rainbow IIIc & 710600 & 156 & 3/4 \\
            Rainbow Vc & 1683300 & 204 & 5 \\
            \hline
            Dilithium $ 1024\times 768 $ & 1184 & 2044 & 1 \\
            Dilithium $ 1280\times1024 $ & 1472 & 2701 & 2 \\
            Dilithium $ 1760\times1280 $ & 1760 & 3366 & 3 \\
            \hline
            FALCON-512 & 897 & 657 & 1 \\
            FALCON-768 & 1441 & 993 & 2/3 \\
            FALCON-1024  & 1793 & 1273 & 4/5 \\
            \hline
            qTESLA-p-I & 14880 & 2592 & 1 \\
            qTESLA-p-III & 38432 & 5664 & 3 \\
            \hline
            Picnic-L1-UR & 32 & 53961 & 1 \\
            Picnic-L3-UR & 48 & 121845 & 3 \\
            Picnic-L5-UR & 64 & 209506 & 5 \\
            \hline 
            SPHINCS$^+$-128s & 32 & 8080 & 1\\
            SPHINCS$^+$-192s & 48 & 17064 & 3\\
            SPHINCS$^+$-256s & 64 & 29792 & 5
            
        \end{tabular}
        \caption{Comparison of the key and signature sizes of the NIST round 2 signature submissions. All sizes given to the nearest byte.}
        \label{fig:postq Key and sig comp table}
    \end{table}
    \noindent
    The timings, however, do have a certain dependence on implementation and as such NIST have set out their requirements with respect to the \emph{NIST PQC reference platform} \cite{NISTcall}: an Intel x64 running Windows or Linux and supporting the GCC compiler. A comparison of the performance of the timings of the schemes can be found in table \ref{fig:Sig and ver tiems comp table}. Unless noted otherwise, the submissions used the reference architecture.
    
    \begin{table}[h!]
        \centering
        \begin{tabular}{c||r|r|r}
            Submission & Key Gen & Signing & Verification \\
            \hline
            \hline
            GeMSS 128 & 38500 & 750000 & 82 \\
            GeMSS 192 & 175000 & 2320000 & 239 \\
            GeMSS 256 & 532000 & 3640000 & 566 \\
            \hline
            MQDSS 48 & 1142 & 36555 & 26639 \\
            MQDSS 64 & 2671 & 116772 & 84685 \\
            \hline
            LUOV-8-58-237 & 17000 & 5400 & 4300 \\
            LUOV-8-82-323 & 71000 & 15000 & 1100 \\
            LUOV-8-107-371 & 127000 & 24000 & 18000 \\
            \hline
            Rainbow Ia & 35000 & 402 & 155 \\
            Rainbow IIIc & 340000 & 1700 & 1640 \\
            Rainbow Vc & 757000 & 3640 & 239 \\
            \hline
            Dilithium $ 1024\times768 $ & 243 & 1058 & 273 \\
            Dilithium $ 1280\times1024 $ & 371 & 1562 & 376 \\
            Dilithium $ 1536\times1280 $ & 471 & 1420 & 511 \\
            \hline
            FALCON 512 & 6.98* & 6081.9$^\dagger$ & 37175.3$^\ddagger$ \\
            FALCON 768 & 12.69* & 3547.9$^\dagger$ & 20637.7$^\ddagger$ \\
            FALCON 1024 & 19.64* & 3072.5$^\dagger$ & 17697.4$^\ddagger$ \\
            \hline
            qTESLA p-I & 2359 & 2299 & 814 \\
            qTESLA p-III & 13151 & 5212 & 2102 \\
            \hline
            Picnic-L1-UR & 160 & 172560 & 116494 \\
            Picnic-L3-UR & 392 & 549036 & 368492 \\
            Picnic-L5-UR & 753 & 1234713 & 828446 \\
            \hline
            SPHINCS$ ^+ $ 128s simple & 326805 & 4868849 & 5304 \\
            SPHINCS$ ^+ $ 192s simple & 486773 & 10259965 & 7971 \\
            SPHINCS$ ^+ $ 256s simple & 636421 & 7570079 & 10866

        \end{tabular}
        \caption{Comparison of the signature creation and verification times of the NIST round 2 signature submissions. Unless stated otherwise these are all to the nearest thousand processor cycles. *milliseconds, $ \dagger $ signatures/second, $ \ddagger $ verifications/second.}
        \label{fig:Sig and ver tiems comp table}
    \end{table}
    
\subsection{Concluding Remarks on Post-Quantum Digital-Signatures}
    Whilst progress in the field of quantum computing does pose a threat to our current digital security models and implementations of digital signature schemes, throughout this section we have given a brief overview of the work being done to combat this direct threat. Working towards NIST’s criteria for both security and efficiency ensures that sought-after solutions are implementable with current technology, well ahead of implementations of Shor’s algorithm being practical. Where theoretical protocols have struggled to reach a compromise between security and efficiency, research has already yielded results to adapt, as shown by modifications of the UOV-schemes and SPHINCS.\\
     
    Indeed, this is the case in reality also, with Google implementing a lattice-based protocol in their Chrome web-browser (although this has since been removed in an effort to ‘not influence the standardization procedure’)\cite{GooglePQC}. However, we remain wary of further threats presented. For example, that a solution to the HSP problem could cause issues for lattice cryptography showcases the need for research to continue remaining ahead of the curve. \\
     
    The National Institute for Standards and Technology's goal of implementing a new standard is predicted to still be at least a year off completion so it is far from finalised. It is certainly likely that the recommendation will be several new standards depending on the application. Whilst there is a notion of simplicity leading to security, some of the above schemes eschew this in favour of ruthless efficiency (albeit with the necessity for careful implementation), with some even claiming to be faster than current protocols. Ultimately the advances in cryptography in response to quantum computing appear to be leading to altogether more complicated systems, but for the sake of security this is certainly the right move.
\section{Quantum Digital Signatures}
\subsection{Key Concepts General to QDS\label{sec: QDS Key Concepts}}

The security of classical digital signatures lies in creating problems that are infeasible to solve. The security of techniques based on quantum physics instead relies upon proven scientific principles\cite{Xu19}. The uncertainty inherent in quantum physics has in recent years found a great many uses in the field of security, ranging from random number generators to optical identity tags\cite{McGrath19}. This well known phenomena is what protects a system against attack, as in many cases, it is physically impossible for the attacker to breach the system without detection.\\
 
As with many classical digital signatures, quantum digital signatures rely upon a one way function for their encryption. In this case however, rather than using a mathematical one way function, classical data is encoded as quantum information\cite{Gottesman01}. In order to implement this each quantum digital signature protocol follows three steps similar to those in classical cryptography\cite{AdvInQC}: 

\begin{enumerate}
\item GEN: Alice uses her private key to generate a signature $s$ consisting of quantum information. 
\item SIGN: Alice sends her message $m$ to the recipients (denoted Bob and Charlie) with the corresponding signature, denoted as $(m, pk)$.
\item VER: Bob and Charlie verify the message is authentic and repudiation has not occurred. In the general case this involves a comparison of $(m, s)$ to the classical description of $s$.
\end{enumerate}

In order to delve further into what each step entails we will discuss a generic QDS model, although schemes will vary in their specific implementation of these steps most follow this general framework\cite{AdvInQC}\cite{Gottesman01}.\\
 
The generation step begins with a purely classical operation, the random generation of a private key for each possible single bit message (1 or 0). This key, denoted as $pk^{i} = (pk^{i}_{1},pk^{i}_{2},...,pk^{i}_{L})$, is purely classical information, where $i =0,1$ to denote the message. Its length $L$ is determined by the level of security required and the QDS scheme used. It is using this string that Alice will identify herself at a later stage so it is imperative it is never shared.\\
 
The next stage of the generation step is done by first defining a set of non-orthogonal quantum states\cite{AdvInQC}. An example of quantum states that can be used is the BB84 states\cite{Singh14}. Alice then generates four separate strings of quantum information (known as quantum digital signatures) by encoding her $pk$ strings using the defined quantum states. These four signatures consist of a copy of the encoded  private key for both possible messages for both Bob and Charlie. These are denoted as $qs^{i}_{B}, qs^{i}_{C}$ with the subscripts denoting who the signature pairs will be sent to. Alice then sends $qs^{i}_{B}, qs^{i}_{C}$ to the correct recipient via a secure quantum channel. Bob and Charlie measure their quantum signature pairs to generate a classical signature from them, $s^{i}_{B}, s^{i}_{C}$ . \\
 
Before proceeding to the next step in most cases Bob and Charlie randomly select approximately half of the elements in their measured signature (though this can occur before measurement) and forward to the other. They do not have to exchange the same elements. As such to all extents and purposes from Alice's perspective both $s^{i}_{B}$ and $s^{i}_{C}$ are exactly the same. This prevents her from committing repudiation. The exact method of this "symmetrisation" is dependant on the QDS scheme.\\
 
To sign a message in most protocols Alice sends her message of a bit of 1 or 0 with the corresponding private key to Bob\cite{Donaldson16}, denoted $(m^{i},pk^{i})$. As the protocols focus only on the signing of a message it is assumed that the message is sent along a secure channel whether this be quantum or classical in nature\cite{AdvInQC}. To send a multi-bit message the process of generation and signing is iterated for each bit\cite{Wang15}.\\
 
Finally there is the verification stage. This varies greatly between each protocol (see relevant protocol section for specific details). The general case is that Bob compares his measured signature $s^{i}_{B}$  to the private key $pk^{i}$ received from Alice. If the number of mismatches between the two is below the required threshold (see section \ref{sec: Quantifying Authentication}) Bob deems the message as authentic. If Bob wishes to forward the message to Charlie, he sends the $(m^{i},pk^{i})$ he received from Alice. Charlie then performs the same process with a different required threshold. \\
 
The security of quantum digital signatures is shown most prominently in the validation step. Each of the signing protocols relies on the same principles to provide security.  It is key to this that the states chosen are non-orthogonal. Therefore, any measurement performed on one state will not commute with a measurement on the other\cite{Xu19}. Thus, any measurement performed will probabilistically disturb the other state, effectively destroying the information it held\cite{Andersson06}. As such without knowing the initial private key no one can discern the original classical input. Getting the correct result is entirely dependent on chance, even then one would have no way to tell if it is the correct result.\\
 
This is reinforced by the Holevo bound placed on the information that can be obtained, only a single bits worth of information may be extracted from a qubit\cite{Holevo73}. This prevents further information to help the attacker's deductions from being obtained. If anyone attempted to forge a signature they would have to guess the private key correctly based on the information they can gather. For short private keys this is indeed improbable but still possible. As a signature gets longer however the chance of successfully guessing falls off exponentially to a negligible value for a long enough private key. As such a simple comparison will reveal their ruse\cite{Dunjko14}. Therefore unlike classical digital signatures QDS is not dependant on the difficulty of mathematical techniques and as such demonstrates information theoretic security\cite{Xu19}. 

%\begin{figure}[H]
%  
%   \centering
%   \captionsetup{justification=centering}
%   \includegraphics[width=6in]{PolarisationExample.png}
%   \caption{Displayed on the left is the two possible polarisation bases for a photon used commonly in encoding the BB84 states. There are four possible states across the two bases. Vertical, horizontal, 45-degree and 135-degree (both from the vertical), the first two make up the rectilinear basis and the latter two the diagonal. Vertical and 45-degree are assigned the value of 1, horizontal and 135-degree the value of 0. Measurements performed in one basis do not commute with those performed in the other. For example, to encode a value of 1, Alice randomly chooses one of the two bases. In the case of measurement 1 she chooses diagonal. When Bob measures the encoded photon he must choose to use one of the bases without knowing what Alice chose. If he chooses diagonal he will get the correct value of 1 from measuring the photon in the 45-degree state. If he chooses rectilinear however, he will get 1 or 0 with equal probability. Similarly if Alice encodes 0 in the rectilinear basis (measurement 2), Bob will only get the correct answer if he measures in the same. Without knowledge of Alice's starting parameters therefore, he cannot with confidence determine Alice's bit. Thus, demonstrating a quantum one way function. }.
%   \label{fig: PolarisationExample}
%\end{figure}

\subsubsection{Quantifying Authentication \label{sec: Quantifying Authentication}}

There are three possible levels to the degree of verification that Bob and Charlie can deem the message/signature combination has fulfilled \cite{Gottesman01}:

\begin{itemize}
\item \textit{1-ACC}: Message is valid, can be transferred.
\item \textit{0-ACC}: Message is valid, might not be transferable.
\item \textit{REJ}: Message is invalid.
\end{itemize}

1-ACC and 0-ACC provide similar levels of security in the standards they uphold, both require that the message be valid. As such if Bob performs the validation test on the signature and finds it to be true then the first condition of these security levels are fulfilled. The difference arises in the transferability of the signature. If Bob has any reason to believe that Charlie would not come to the same conclusion as him then the message and signature would be deemed 0-ACC (i.e. repudiation has occurred). Only a message with validation level of 1-ACC is deemed fully secure. Finally if the signature that Bob receives is invalid then the message is given the validation level REJ and is rejected. \\
 
These criteria are mathematically defined by a series of thresholds proposed by Gottesman and Chuang\cite{Gottesman01}. We first define the probability of failure of an attacker $p_{f}$, this is the probability their measurement will fail to get the correct classical description given a minimum error measurement. There is also the probability that an honest measurement will fail due to environmental factors such as noise, denoted $p_{e}$. Both of these thresholds are calculated based on outside factors such as the number of copies of the signature circulated, the method with which measurements are taken, the apparatus used to distribute/measure the signatures, etc. Using these as the boundaries for acceptable error levels, with $p_{f}$ as the upper and $p_{e}$ as the lower, the allowed error thresholds can be set\cite{Gottesman01}. For a signature to be authenticated it must have a fractional error lower than $s_{v}$, wherein $p_{f} > s_{v}$. Based on this the upper bounds of the probability that the forger can effectively mimic a valid signature is given by: $e^{-c(p_{f}-s_{v})^{2}L}$\cite{AdvInQC}, where $c$ is a constant. This shows that the security against forging is dependant not only on the environmental factors (as $p_{e}$ and $p_{f}$ define the range in which $s_{v}$ can fall) but on the length of the signature itself\cite{AdvInQC}.\\
 
One threshold is not enough to protect against repudiation as stated by Gottesman and Chuang\cite{Gottesman01}. If Alice was dishonest and sent different signatures to Bob and Charlie (with the aim to have Bob accept and Charlie reject it), she could successfully repudiate with only one threshold. With only one threshold ($s_{v}$) for verifying the signature Alice could tailor each signature to have $s_{v}L$ errors. The number of errors is defined by a normal distribution with a mean of $s_{v}L$. As such in the probability of Bob accepting the message (errors below $s_{v}L$) and Charlie rejecting it (errors above $s_{v}L$) is 0.25\cite{AdvInQC}. By introducing a second threshold $s_{a}$ where $s_{v} > s_{a}$ that Charlie must pass instead of $s_{v}$ reduces the probability of successful repudiation to negligible with a long enough signature. This is due to the fact that Alice would have to generate a signature that would give a result both below $s_{v}$ and above $s_{a}$. In a similar manner to forgery the upper bound on the probability for successful repudiation is now defined by $e^{-c'(s_{a}-s_{v})^{2}L}$, where $c'$ is a constant. \\
 
This defines the necessary criteria for setting out mathematically how to achieve the verification conditions. To summarise the relative size of each of the thresholds: 0$<p_{e}<s_{a}<s_{v}<p_{f}$. In order to achieve a 1-ACC level of verification a signature's error fraction must be below $s_{v}$ for Bob and $s_{a}$ for Charlie. Falling below only $s_{v}$ would result in a 0-ACC rating and falling below neither results in a REJ.\\

\subsection{QDS with Quantum Memory \label{sec: QDSwithQM}}

The first QDS protocol was proposed by Gottesman and Chuang in 2001\cite{Gottesman01}, hence referred to as GC-QDS. As the precursor to all other QDS protocols as expected GC-QDS was only a theoretical proposal, however it is important to analyse. It sets the standard for the "ideal" quantum digital signature protocol. One in which the signature remains as quantum information throughout the entire process. This ensures information theoretic security throughout. \\
 
As detailed in section \ref{sec: QDS Key Concepts}, to begin with Alice generates her random private keys, $pk^{i}$, for each possible message value. The initial proposal for this scheme suggested each private key element, $pk_{n}^{i}$, be a two bit string and $qs_{n}^{i}$ the corresponding BB84 state\cite{Gottesman01}. Using this, Alice generates copies of each quantum digital signature for each possible message value by encoding $pk^{i}$ as quantum information. As many copies of the public key as there are recipients are generated and one distributed to each. The recipients, in this case Bob and Charlie, then do not measure the received signature and instead store it in stable quantum memory, a key difference to other protocols. This is the "generation" phase for this protocol\cite{Gottesman01}.\\
 
For the signing phase, Alice only sends out classical information. She sends to Bob $(m^{i},pk^{i})$. As with most QDS protocols this is assumed to be performed over a secure classical channel. An assumption that can be inexpensively implemented and so is not outlandish to assume. \\
 
Finally there is the verification phase. Using $pk$ and the known function for encoding classical information as quantum information Bob generates his own set of quantum states. He then compares the states he has generated to those he has stored from Alice using a SWAP test\cite{Gottesman01}. If the number of mismatches between the two falls below the acceptance threshold $s_{a}$, Bob accepts the message as authentic. For further verficiation he can forward $(m^{i},pk^{i})$ to Charlie who will repeat the process with his threshold $s_{v}$. \\
 
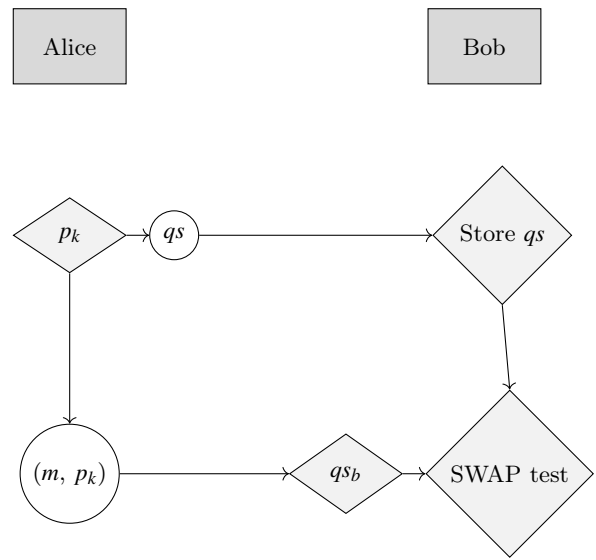
\begin{figure}
        \begin{center}
        \begin{tikzpicture}
            %\tikzset{}
            
            % Nodes
            
            \node[PrivInfo]             (PK)          {$p_{k}$};
            \node[Party]  (Alice)            [above=1.5cm of PK]                 {Alice};
            \node[Party]  (Bob)          [right=4cm of Alice]                   {Bob};
            \node[PubInfo]             (Sign)       [below=2cmof PK]           {($m$, $p_{k}$)};
            %\node[Protoblock]             (Public)      [right=0.5cm of PK,text width=2cm]      {Create public key from private};
            \node[PubInfo]             (Send)[right=0.3cm of PK]             {$qs$};
            \node[PrivInfo]  (Storage)           [right=3.1cm of Send]        {Store $qs$};
            %\node[PubInfo]             (SendM)[right=0.5cm of Sign,text width=3cm]             {Send $(m, p_{k})$ to Bob};
            \node[PrivInfo]             (Ver1)[right=2.25cm of Sign]             {$qs_{b}$};
            \node[PrivInfo]             (Ver2)[right=0.3cm of Ver1]             {SWAP test};
            %\node[sqarednode]   (checking)          [above right=of hashed, yshift=-1.3cm]{};
        
            % Lines
            \draw[->]   (PK.east) -- (Send.west);
            %\draw[->]   (Public.east) -- (Send.west)
            \draw[->]   (Sign.east) -- (Ver1.west);
            \draw[->]   (Send.east) -- (Storage.west);
            \draw[->]   (PK.south) -- (Sign.north);
            %\draw[->]   (SendM.east) -- (Ver1.west)
            \draw[->]   (Ver1.east) -- (Ver2.west);
            \draw[->]   (Storage.south) -- (Ver2.north);
            %\draw [dashed, ultra thick] (8,3) -- (8,-6);
           % \draw[<->]  (hashed.east) -- (secondverification.west)
        \end{tikzpicture}
        \end{center}
        \caption{Flow diagram breaking down the process of Gottesman and Chuang's first proposal for a quantum digital signature protocol. Alice begins by randomly generating her private key ($p_{k}$). From this she generates her quantum signature ($qs$) which is sent on to Bob. Bob stores this in quantum memory without measuring it. To sign a message ($m$), Alice sends it alongisde its corresponding $p_{k}$ to Bob. Bob uses $p_{k}$ to genreate his own copy of the quantum signature $qs_{b}$. He verifies its authenticity in a SWAP test with $qs$. Diamonds represent information that must be kept private (at least until sending), circles represent information that is sent to another individual. In the case of this QDS scheme the public key is quantum information, all other information is classical.}
        \label{fig: QDSDiagramQM}
    \end{figure}
    
\noindent The unforgeability in this scheme stems from the strict policy of not measuring the quantum digital signature until authentication is required\cite{AdvInQC}, as demonstrated in figure \ref{fig: QDSDiagramQM}. For an attacker to forge a signature they would as in, classical digital signatures, have to bypass the one-way nature of of the classical to quantum encoding. Aside from obtaining the private key from Alice the only way to achieve this is to intercept the quantum digital signature and correctly guess the measurement basis for each qubit. Thus, as previously discussed, for a long enough signature there is a negligible probability that Bob and Charlie will not notice that the signature has been tampered with\cite{Dunjko14}. Owing to the collapse of a quantum wavefunction upon measurement, the very act of attempting to intercept and forge a message will reveal an attack has occurred as the signature remains only as quantum information. The only attack that can be achieved by interception is to cause the $qs$ distribution phase to abort \cite{Gottesman01}.\\

The protection against repudiation lies with the comparative SWAP tests and relevant thresholds that Bob and Charlie apply to their stored public keys\cite{Gottesman01}. As SWAP tests do not measure a quantum state and instead compare two to determine similarity it makes them the perfect operation to enforce non-repudiation. Coupling the non-destructive nature of the SWAP test with the thresholds detailed in section \ref{sec: QDS Key Concepts} renders the probability of repudiation to negligible with a long enough signature\cite{Gottesman01}.\\

This scheme is not without faults however, its most prominent fault is its reliance on the immature technology of quantum memory. If the technology were perfected it would allow for the indefinite storage of $qs$. As of time of writing however, quantum memory can not store quantum information for long time periods \cite{Bouillard19}\cite{Collins14}. In theory this protocol can have long term quantum digital signatures but in practice this is simply not possible. Although there have been recent advancements in the storage times of quantum memories this protocol is currently infeasible\cite{Donaldson16}\cite{Roberts17}.\\

The SWAP tests themselves are an issue within the same vein as quantum memory, the technology to perform them is not available. Each recipient would require a quantum computer in order to perform such a test. As with the quantum memory requirement this ensures Gottesman and Chuang's theoretical proposal remains theory.

%finsih the basic steps then move onto why it is secure, interception and repuftion!  then add references and diagram and done!

%"Quantum cryptography with realistic devices" HAS A ReVIEW OF METHODS IMPLEMENTED UP TO 2019!!

\subsection{Multiport set-up\label{sec: Multiport}}
%Specifically in to raise to light some of the issues in QDS schemes such as the noise (can use the data in paper for that??), proximity of bob and charlie, complex equipment, simultaneous measurements etc... Excellent spot to critque the literature 

The attractive concept of quantum digital signatures coupled with the initial proposal's reliance on quantum memory and computing lead to a great deal of interest in QDS from both a theoretical and experimental standpoint\cite{Xu19}. The reliance on currently immature quantum technologies has led to new proposals that find methods of getting around this constraint. One of the earliest proposals was that of the multiport \cite{Clarke12}.  This apparatus consists of a square array of four separate 50:50 beam splitters as shown in figure \ref{sec: Multiport}\cite{Dunjko14}. The  apparatus and its potential in cryptography was first proposed in 2006 \cite{Andersson06}but only as a theoretical method of public key distribution. Later, in 2012, it was adapted for use in quantum digital signatures and experimentally demonstrated.\\

For use in quantum digital signatures the multiport is effectively split in two, the top two beam splitters belong to Bob and the bottom two to Charlie. The multiport in of itself primarily affects the generation stage of the quantum digital signature process. \\

The generation stage begins the same as other QDS schemes. Alice generates a randomised classical private key which she encodes with her chosen quantum basis. She then sends these to Bob and Charlie. She sends the the copies of each quantum digital signature at the same time (i.e. $qs^{1}_{B}$ and $qs^{1}_{C}$ are sent out at the same time and then the other set of copies are sent).\\

\begin{figure*}[ht]
        \begin{center}
        \begin{tikzpicture}[
            ]
           \node[draw, diamond, line width=0.50mm, scale =6] (multiport1) {};
           \node[draw, diamond, line width=0.50mm, scale =6] (multiport2) [right =3cm of multiport1] {};
           \node[draw, diamond, line width=0.50mm, scale =6] (multiport3) [below =3cm of multiport1]{};
           \node[draw, diamond, line width=0.50mm, scale =6] (multiport4) [right =3cm of multiport3] {};
           \node[] (start1) [below left =0.8cm of multiport1, xshift=0.1cm ] { $\ket{\alpha}$};
           \node[] (start2) [above left =0.8cm of multiport3, xshift=0.1cm ] { $\ket{\beta}$};
           \node[] (end2) [below right =1cm of multiport2, yshift = 0.0cm, xshift=0.3cm] { $\ket{ \frac{\alpha - \beta}{2}}$};
           \node[] (end2b) [above right =1cm of multiport4, yshift = 0.0cm, xshift=0.3cm] { $\ket{ \frac{\alpha - \beta}{2}}$};
           \node[] (end1) [above right =1cm of multiport2, yshift = 0.0cm, xshift=0.3cm] { $\ket{ \frac{\alpha + \beta}{2}}$};
           \node[] (end3) [below right =1cm of multiport4, xshift=0.3cm] { $\ket{ \frac{\alpha + \beta}{2}}$};
           \node[Party]  (Bob)            [right =2cm of multiport2]                 {Bob};
           \node[Party]  (Charlie)            [right =2cm of multiport4]                 {Charlie};

           \draw[-, line width=0.50mm]   (multiport1.east) -- (multiport1.west);
           \draw[-, line width=0.50mm]   (multiport2.east) -- (multiport2.west);
           \draw[-, line width=0.50mm]   (multiport3.east) -- (multiport3.west);
           \draw[-, line width=0.50mm]   (multiport4.east) -- (multiport4.west);
           \draw[blue, -]   (multiport1.center) -- (multiport4.center);
           \draw[red, -]   (multiport3.center) -- (multiport2.center);
           \draw[blue, -]   ([yshift=-0.1cm, xshift=0.1cm]start1.north) -- (multiport1.center);
           \draw[red, -]   ([yshift=0.1cm, xshift=0.1cm]start2.south) -- (multiport3.center);
           \draw[-, line width=0.60mm] (1,3)--(5,3);
           \draw[-, line width=0.60mm] (1,-9)--(5,-9);
           \draw[blue, -]   (multiport1.center) -- (3,3);
           \draw[blue, -]   (3,3) -- (multiport2.center);
           \draw[red, -]   (multiport3.center) -- (3,-9);
           \draw[red, -]   (3,-9) -- (multiport4.center);
           \draw[violet, ->]   (multiport2.center) -- (end2.west);
           \draw[violet, ->]   (multiport4.center) -- (end2b.west);
           \draw[violet, ->]   (multiport2.center) -- (end1.west);
           \draw[violet, ->]   (multiport4.center) -- (end3.west);
        \end{tikzpicture}
        \end{center}
        \caption{Schematic diagram of the multiport set-up used in the multiport QDS scheme. Each bisected diamond represents a beamsplitter. The thick black lines at the top and bottom of the diagram represent mirrors. One half of the array is in the posssesion of Bob and the other in the possession of Charlie.}
        \label{fig: Multiport}
    \end{figure*}
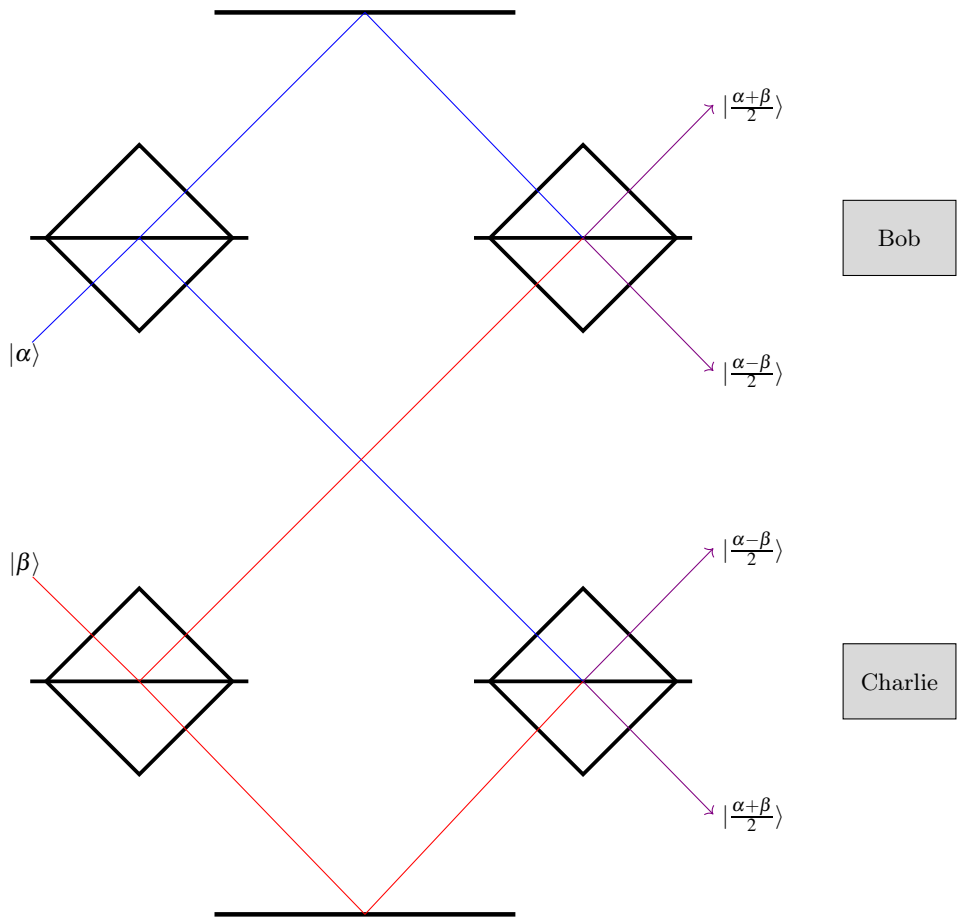

The first set of beam splitters (moving left to right on figure \ref{fig: Multiport}) are used by Bob and Charlie to split their copies of the signature into two equal amplitude components. One half of these are kept by Bob/Charlie and the other half are sent to the other recipient. This ensures that Alice is unaware as to who has which bit in each of the copies that she initially sent\cite{Dunjko14}\cite{Collins14}. At the second set of beam splitters the half that was originally kept by the receiver is mixed with the half from the other recipient in a comparison test. The process of this is detailed in figure \ref{fig: SingleMultiport}. \\

\begin{figure}[h!]
        \begin{center}
        \begin{tikzpicture}
           \node[draw, diamond, line width=0.50mm, scale =10] (multiport1) {};
           \node[] (alpha) [above left = 0.5cm of multiport1, xshift=-0.2cm] { $\ket{\alpha}$};
           \node[] (beta)[below left = 0.5cm of multiport1, xshift=-0.2cm] { $\ket{\beta}$};
           \node[] (end1)[above right = 0.5cm of multiport1, xshift=0.3cm, yshift=-0.1cm] { $\ket{\frac{\alpha + \beta}{2}}$};
           \node[] (end2)[below right = 0.5cm of multiport1,xshift=0.3cm, yshift=0.1cm] { $\ket{\frac{\alpha - \beta}{2}}$};

           \draw[-, line width=0.50mm]   (multiport1.east) -- (multiport1.west);
           \draw[blue, -]   (alpha.east) -- (multiport1.center);
           \draw[red, -]   (beta.east) -- (multiport1.center);
           \draw[violet, ->]   (multiport1.center) -- (end1.west);
           \draw[violet, ->]   (multiport1.center) -- (end2.west);
           
        \end{tikzpicture}
        \end{center}
        \caption{Schematic diagram of a single beamsplitter. From the left enters two separate photon packets, $\ket{\alpha}$ and $\ket{\beta}$. These are combined by the splitter to give  $\ket{\frac{\alpha + \beta}{2}}$ and $\ket{\frac{\alpha - \beta}{2}}$.}
        \label{fig: SingleMultiport}
    \end{figure}
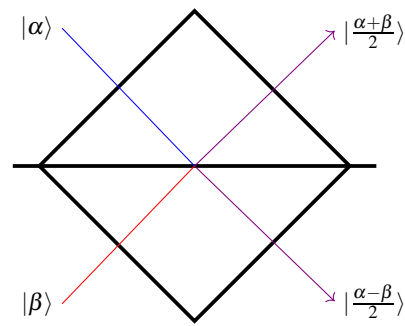

\noindent In the initial implementation of this protocol, Bob and Charlie would store the received states in quantum memory ready for the verification stage. In 2014 Dunjko showed that it was indeed possible to remove the quantum memory from this protocol\cite{Dunjko14}. Progressing instead to verification via comparison of classical data, the latter will be the focus of this section.\\

Rather than storing the quantum digital signature, the incoming signature is measured and the outputs of this measurement stored. This classical string of results then forms an individual's measured signature. Most commonly the measurements performed are quantum state elimination measurements (covered in further detail in section \ref{sec: RandomForwarding}), however unambiguous state discrimination was initially proposed\cite{Collins14}. The key principle of either method is that it does not give a completely accurate description of Alice's initial private key. Thus, it does not enable a recipient to then forge a signature. The measurement and storage of this now classical signature completes the generation stage of the protocol. \\

To sign a message Alice simply sends her single bit message alongside the relevant private key to Bob\cite{Collins14}. Bob then compares the private key to the signature he measured, applying the $s_{a}$ threshold detailed in section \ref{sec: Quantifying Authentication} to determine the degree to which he trusts it. He then forwards the key and message to Charlie to validate that Alice indeed sent them both the same signature and similarly Charlie compares the two to see if the errors between them fall below $s_{v}$. From these results they determine the level of validity in the message and signature. \\

The security of this protocol lies with the square array of beam splitters\cite{Clarke12}. It is the secondary beam splitters that Bob and Charlie use to check the validity of the signature. One output of the beam splitters will be a mixed state of $\ket{\alpha}$ and $\ket{\beta}$. If Alice was honest then  $\ket{\alpha}=\ket{\beta}$ and as both are identical and coherent the initial input from Alice is obtained. If, however, Alice sent Bob and Charlie different signatures the multiport will symmetrize these and prevent repudiation\cite{Clarke12}. The null ports then act as a safeguard against active forging\cite{Collins14}. In this scenario Bob is the dishonest party and attempts to forward a forged signature and message to Charlie. In this case Charlie would measure a non-zero (assuming no background count) reading on  his null port, informing him to the presence of a forged signature. \\
One of the key issues with this scheme is the loss present in this system, measured at 7.5dB. The signature length $L$ required for a security level of 0.01$\% $ was $5.0\times 10^{13}$ for a half bit\cite{Collins14}. The count rate with USE was found to be $2.0\times 10^{5}$ counts per second. Given that this would yield a time required of 7.9 years to sign and send this is clearly an impractical signature length. This is particularly troublesome when one considers this was not taken at a separation distance, $5m$, great enough for use in a practical setting\cite{Collins14}. Methods of improving upon this technique were proposed such as increasing the clock rate of the VCSEL used to generate the pulses. Due to the loss rates and impractical distance requirements however, using multiports in quantum digital signatures serves only as a proof of concept for not requiring advanced quantum technologies.
\subsection{Random Forwarding\label{sec: RandomForwarding}}
%Detail BB84 states. Remember to cover if the different methods use USE! Two variation, forward the quantum states and forward the measurements of them

The development of quantum digital signature schemes shows common themes, the reliance on quantum mechanics to achieve information theoretic security and moving away from the reliance on immature quantum technologies. Multiports, whilst flawed, demonstrated that quantum memory and quantum computing is not necessary for QDS. The next clear step, as stated in the paper implementing multiports, is to develop a system that does not use them for security\cite{Collins14}. The simplest way to achieve this is random forwarding. \\

\begin{figure}[h!]
        \begin{center}
        \begin{tikzpicture}[
            squarednode/.style={rectangle, draw=black!60, thin, minimumsize=5mm},
            ]
           \node[] (a)[] { Alice};
           \node[] (b) [above right = 3.0cm of a, xshift=0.20cm] { Bob};
           \node[] (c) [below right = 3.0cm of a] { Charlie};
           \node[] (1) [above right =0.5cm of a, xshift =0.7cm] {1};
           \node[] (4) [above right = 1.5cm of a, xshift =-.55cm] {4};
           \node[] (2) [below right = 1.5cm of a, xshift =-0.50cm] {2};
           \node[] (3) [below = 2cm of b, xshift =0.2cm] {3};
           \node[] (5) [below = 2cm of b, xshift =-0.35cm] {5};
           
           \draw[blue, ->]   (a.east) -- ([yshift=-0.2cm]b.west);
           \draw[blue, ->]   (a.east) -- (c.west);
           \draw[blue, <->]   (c.north) -- (b.south);
           \draw[red, ->]   ([xshift=0.45cm, yshift=0.03cm]a.north) -- ([xshift=-0.1cm, yshift=0.00cm]b.west);
           \draw[red, ->]   ([xshift=-.2cm]b.south) -- ([xshift=-.2cm]c.north);
           
        \end{tikzpicture}
        \end{center}
        \caption{Representation of the communication channels between Alice, Bob and Charlie in QDS schemes based on random forwarding. Blue arrows represent quantum communication channels and red classical channels. Single headed arrows represent one way communication channels, double headed represent two way channels.}
        \label{fig: RandomForwardingLayout}
    \end{figure}
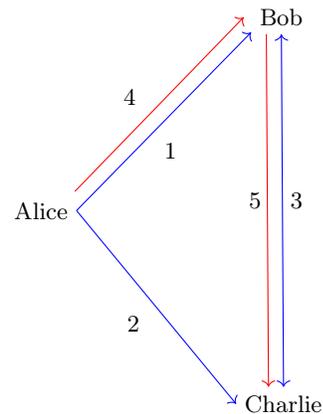

As with all previous schemes, Alice begins by generating a string of classical bits that she keeps as her private key. She then encodes this in quantum states using non-orthognal bases\cite{Wallden15}. Once again two copies of each quantum digital signature are created and the copies of the same signature are sent to Bob and Charlie at the same time (arrows 1 and 2 on figure \ref{fig: RandomForwardingLayout}). \\

Unique to this protocol is that upon receiving the quantum digital signature Bob and Charlie randomly choose elements of the signature $qs^{i}$ to forward to the other, usually by a "coin toss" protocol\cite{Wallden15}\cite{Donaldson16}, arrow 3 on figure \ref{fig: RandomForwardingLayout}. They then record the location in the string of those that were passed on and which elements were retained. If either receive less than $L(\frac{1}{2}-r)$ or more than $L(\frac{1}{2}+r)$, where $r$ is a threshold the two of them set, they abort\cite{Wallden15}. Thus Bob and Charlie's final quantum digital signature will be a randomised mix, who's contents is defined only by the "coin toss" performed to dictate whether to keep an element. From the view point of Alice therefore once she has sent the messages the reduced density matrices for Bob’s and Charlie’s quantum digital signature elements are identical, regardless of whether or not she tried to commit repudiation\cite{Wallden15}.\\

Bob and Charlie then measure their quantum signature copies to get their measured classical signature $s^{i}$. This is known as the pre-measurement approach\cite{Wallden15}. Alternatively Bob and Charlie can measure the quantum digital signature elements before forwarding in a post-measurement approach. In this case there is no need for a quantum communication channel between them, reducing the system to only having the quantum channels between them and Alice. In either scheme the technique of USE (Unambiguous State Elimination) is commonly used to measure $qs$\cite{Donaldson16}.  \\

To sign a message Alice simply sends her private key concatenated with the message to Bob (arrow 4 on figure \ref{fig: RandomForwardingLayout}). Bob verifies the authenticity of the signature by comparing how many signature elements he correctly eliminated. If the error in this falls below $s_{a}$ then the message is deemed authentic and he passes it along with the key to Charlie. Charlie performs the same comparison with his stored classical signature and his error threshold $s_{v}$. If it passes this then the message is deemed valid. Repudiation has not occurred due to the symmetrised signatures and Bob has not committed a forgery as Charlie's signature has successfully been compared with Alice's.\\

\noindent The benefits of random forwarding are apparent in the reduced dependence on quantum technologies. Stripping back QDS protocols so one is only using QKD to handle quantum information is far more practical than quantum memory or a multiport\cite{Donaldson16}\cite{Collins16}. The technology is more mature and has undergone a rigorous amount of field testing. Thus, allowing for easier integration of QDS into existing networks. In addition using QKD based quantum communication adds, in exactly the same manner for QKD, protection against message interception. Although this scheme would result in the sacrifice of some bits for Alice and Bob to compare it would remove the risk of outside interception\cite{Amiri16}. Giving the random forwarding protocol a wide range of applications and versatility. As such many other schemes build on the primitive of random forwarding. Either by developing more advanced hardware and measurement techniques\cite{Croal16} or using it as the primitive for symmetrisation in schemes such as a QKD based schemes proposed by Collins et al. \cite{Collins16} detailed in section \ref{sec: QKDBasedSignatures}. 

\subsection{QKD Key Generation Protocol\label{sec: QKDBasedSignatures}}

Quantum cryptography as a field did not begin with the development of QDS, but instead with a far more developed technique is that known as Quantum Key Distribution. This is a process by which, through the exchange of quantum information, Alice and Bob can generate a secret key for use in encrypted communication. By observing the error rates that occur in the measurement of the quantum information it is possible to detect if an eavesdropper is present\cite{Xu19}. Through this it can be confirmed if the exchanged key is indeed completely secret. If so by using a one time pad protocol Alice and Bob can achieve completely secure encryption. This in of itself could be used to generate a digital signature\cite{Wallden15}\cite{Amiri16}.  As Alice and Bob would know that the signature could only be known by one of them it would act as an identifier. \\

This principle in of itself however, is not of particular use. Firstly it does not allow for more than one recipient, a major drawback for a signing scheme. Secondly it was shown that schemes based on partial QKD protocols could be expanded to multiple recipients and were more efficient in signing than secret key exchange via QKD\cite{Amiri16}.\\

The Quantum Key Distribution Key Generation Protocol (QKD KGP) differs from other quantum digital signature schemes in how the quantum information is distributed. It does, however, still follow the usual three step process of generation, signing and verification. \\

In the generation step, rather than  Alice sending quantum information to Bob and Charlie, they send it to Alice\cite{Collins16}. This is done to simplify security analysis as it means Alice cannot send out entangled states. As with Alice in other schemes Bob and Charlie first generate their own individual random classical private key for both possible message values ($pk^{0}_{B}$, $pk^{1}_{B}$ and $pk^{0}_{C}$, $pk^{1}_{C}$), recording the basis that each element was encoded in. They encode this as quantum information using the chosen method (both BB84 states and phase encoding have been used\cite{Collins16}\cite{Amiri16}\cite{Yin17}). Thus, forming two sets of separate quantum digital signatures $qs^{0}_{B}$, $qs^{1}_{B}$ and $qs^{0}_{C}$, $qs^{1}_{C}$.\\

Bob (Charlie) then performs a partial QKD protocol with Alice\cite{Amiri16}. He sends his quantum digital signatures to her, Alice chooses a random basis, based on the encoding method used, to measure each element. Resulting in Alice having two strings of classical elements from Bob (Charlie), $s^{0}_{B}$ and $s^{1}_{B}$ ($s^{0}_{C}$ and $s^{1}_{C}$). Bob (Charlie) then announces which basis each element was encoded in for each $qs$ over a classical channel to Alice, who then "sifts" her signatures by discarding any that weren't measured in the same basis. Bob (Charlie) discards the corresponding elements of his private key. Two further sections of the measured signatures and corresponding private keys are then sacrificed\cite{Amiri16}\cite{Yin17}. At first Alice and Bob (Charlie) determine the hamming distance between corresponding sections of their signature and private key. If they are sufficiently correlated (they need not be exactly the same) the process proceeds and those sections are discarded. Secondly, Alice and Bob (Charlie) compare corresponding sections in order to observe if the error that an eavesdropper who induce in the measurements is present. If not these sections are discarded and the process continues. Each test is performed over a classical channel and if either of these steps are not successful the process repeats. \\

Alice will now have 4 sets of signatures, $s^{0}_{B}$, $s^{1}_{B}$, $s^{0}_{C}$ and $s^{1}_{C}$. As the error correction usually present in QKD protocols has not been performed these measured signatures will not exactly match their private key counterparts. Bob and Charlie then randomly select half of their private keys and forward them on to the other over a secret classical channel. To all extents and purposes from the perspective of Alice, each of these keys is identical, as she does not know what has been forwarded\cite{Collins16}. \\

To send a message Alice then sends to Bob $(m^{i}, s^{i}_{B}, s^{i}_{C})$. To verify the authenticity of this, Bob compares $s^{i}_{B}$ to his corresponding private key $pk^{i}_{B}$ and $s^{i}_{C}$ to the half that he received from Charlie. If the fraction of mismatches in both is less than $s_{a}$ then he deems them authentic. For further verification he can forward $(m^{i}, s^{i}_{B}, s^{i}_{C})$ to Charlie who will repeat the process with $s_{v}$ as his threshold. \\

The security of this scheme relies on the already proven security of QKD whilst adapting that pre-existing technology for use as a quantum digital signature\cite{Amiri16}. The key difference to other QDS schemes, that of two different quantum signatures for each message, improves the efficiency of the scheme over both other QDS schemes and secret key sharing via QKD. Neither Bob nor Charlie can be dishonest and attempt to forge as they do not know the half of the other's private key that was not sent. This is opposed to other QDS schemes wherein a forger has access to the whole of the QDS. The different private keys removes the risk of colluding forgers whom in other schemes would have had a copy of the QDS each to try and determine the correct measurement values from it. The only option for a forger is to eavesdrop, which can be determined from the sacrificed bits. As well as this Alice cannot commit repudiation as she does not know who has which private key elements. Finally this scheme's lack of the need for error correction and privacy amplification as in a full QKD protocol means it is more tolerant to noise. As such it can be implemented in QKD based systems and used in a wider variety of scenarios. 

\subsection{Expanding to Signing Multiple Bits\label{sec: MultiBit Signing}}

The protocols detailed in this report have all focused on signing a single bit of data. The message is encoded into the quantum digital signature by having a signature for both possible message values. As of yet there is not a great calling to analyse multi-bit messages as the focus is on producing a practical single bit protocol. It is proposed that single bit signing protocols are expanded in a simple manner to sign a message of many bits\cite{Yao19}.  For each bit in the message as a whole the signing process is iterated. Multiple different signature pairs would be sent to Bob and Charlie. When Alice sends her multi-bit message she would send each bit with a valid private key string. Nonetheless some papers have, however, raised concerns over this. Citing that insufficient research has been performed in this area\cite{Wang15}. As such simply iterating the process may weaken the security of the protocol and not even be the most efficient way to sign the message. 

\subsubsection{Conflict over Protocol Iteration\label{sec: ConflictOverIteration}}

The potential issues surrounding iterating a protocol were first raised by Tian-Yin Wang et al. \cite{Wang15} in which a multi-bit signing scheme was proposed that "tagged" the ends of a message. This was later returned to and improved upon\cite{Wang17}. This work gained attention from other research groups who also saw an issue in single bit iteration. Techniques such as ghost imaging \cite{Yao19}, quantum but commitment\cite{Wang18} and adaptations of Wang's initial technique\cite{Zhang19} have been proposed. \\

The argument for defining how a protocol handles messages longer than a single bit in length arises as the whole multi-bit message itself isn't encoded anywhere in the signature. Only the value of a single bit is encoded. In a classical signature this is the case and allows for the checking of message integrity. A demonstration of the the issues of simply iterating a QDS protocol is that of selective attacks. Defining a message as a series of iterated single bits with corresponding private key strings such that the pair received by Bob ($M, PK_{M}$) can be broken down as\cite{Wang15}: 

\begin{equation}
M = m_{1}||m_{2}||...||m_{n}
\label{eq: MultiBitMess}
\end{equation}
\begin{equation}
PK = pk_{1}||pk_{2}||...||pk_{n}
\label{eq: MultiBitKey}
\end{equation}

Where $m$ and $pk$ represent the individual bit components of the multi-bit message and their corresponding private key strings respectively. The corresponding set of quantum signature strings is therefore denoted by:

\begin{equation}
QS = qs_{1}||qs_{2}||...||qs_{n}
\label{eq: MultiBitSig}
\end{equation}

Where $n$ gives the total bit length of the multi-bit message and $||$ denotes the concatenating of subsequent components. Wang \cite{Wang15} argued that Bob could successfully forge a message by only selectively forwarding certain bit strings. The practical example used in Wang's paper\cite{Wang15} was that $M$ sent by Alice consisted of the bits to represent the phrase "Don't pay Bob \$ 100". As this message was produced by iterating a protocol for each bit, each bit of the message $m_{i}$ would have a corresponding private key string $pk_{i}$. Bob is now in possession of a signed message with a set of associated correct signatures. Wang purported that Bob may simply only forward the bits (and corresponding private key strings) that represent for example the message "Pay Bob \$ 100"\cite{Wang15}\cite{Wang17}. As the private key string with each bit will successfully verify when Charlie checks against his stored quantum signature measurements, Bob has successfully forged a message from Alice.\\

This is not the only attack reported by Wang\cite{Wang15} that could be performed. If Bob has in his possession two separate message and private key sets then, he could separate out the message bits and "stitch" them together in a new order to forge a message. For example if he receives two messages from Alice one stating "Pay Bob \$ 100" and another stating "I have \$ 200", Bob can choose to only forward the bits in the first part of the first message with the latter part of the second to give "Pay Bob \$ 200". As the bit forwarded would have a correct corresponding private key Charlie would verify this message as correct. \\

These would clearly breach the security of a quantum digital signature protocol. The unbreakable security that is derived from the quantum mechanical effects inherent to the system is let down by an issue in how the protocol is implemented. Despite the minimal further research to support Wang's claim the issue raised by it is still valid. Although the examples mentioned were simplistic this could have serious repercussions in real life applications. For example if Alice had sent Bob a contract then he could choose not to send the bits relating to sections of the contract he did not approve of to Charlie. Furthermore these attacks need not even be performed by Bob, if a malicious third party, Eve, were to intercept ($M, PK_{M}$) they could also commit acts of forgery.\\

It is clear that work is needed in expanding protocols to consider how to sign multi-bit messages safely. Although this work is likely not a priority for many research groups until the rapid secure signing of a single bit is developed it is necessary for the full practical implementation of QDS.
\subsubsection{"End Tagging"}

A naive solution to this would be to record the sequence in which the signatures are sent and label the message/private key combinations with the corresponding number\cite{Zhang19}. This would not prevent Bob from omitting information at the end of a message however (unless the number of signatures sent exactly matched the number of message bits) or prevent him from "stitching" together messages. The latter it would only make more difficult as he could take the part of one message (say the first five bits of a ten bit message) and 'stitch' it together with another (the last five bits of the second message). \\

To solve these issues, Wang\cite{Wang15} proposed a protocol to be used as a primitive in an overall multi-bit signing protocol. Any of the single bit signing schemes detailed in this report can be used to sign the individual bits, the "end tagging" process determines how these should be iterated to form a multi-bit message. \\

In keeping with the notation for a multi-bit message given in section \ref{sec: ConflictOverIteration} the first step in the proposed method is to create  a "sufficiently large" \cite{Wang17} number of private key strings. Each of these are labelled as corresponding to a 0 or 1 message bit (in the same manner as with single bit signing protocols) and also sequentially numbered. The method of encoding as quantum information is independent to the rest of the protocol, as per single bit QDS protocols a copy of each signature is generated for each recipient. The distribution of the set of of quantum signatures $S$ to said recipients is unaffected by this multi-bit expansion.  As such any QDS generation and distribution scheme can be used, making this protocol simple to use to extend existing schemes. The recipients then measure each of these to create their own set of signature strings.\\

Where Wang's proposal tackles the issues of multi-bit encoding is in the encoding of the message. The following steps are taken to encode $M$\cite{Wang15}:

\begin{enumerate}
\item Encode any bit with the value of 0 as 00.
\item Encode any bit with the value of 1 as 01.
\item Add the codeword 11 to the start and the end of the message.
\end{enumerate}

This results in an output message $\hat{M}$ which has $2n=4$ elements compared to $M$'s $n$ total elements. To sign a message each bit in the message is assigned a private key string depending on its bit value and on its location in the message (e.g. the first bit will be assigned the first of the private key strings). Alice then sends to Bob the combination of information denoted as $(M, PK_{\hat{M}}, l)$ where the message before encoding is denoted $M$, the private key strings for each bit in the encoded message denoted $PK_{\hat{M}}$ and $l$ the sequence number of the first key. \\

Bob then converts $M$ to $\hat{M}$ in the same manner that Alice did. He then applies the authentication method associated with the method of encoding used to each measured string $s_{i}$ present in the set $S$.  If each string passes authentication then he knows that the message is from Alice and that it has not been tampered with. For secondary verification Bob can forward $(M, PK_{\hat{M}}, l)$ on to Charlie for him to authenticate. \\

The security of the signing of each individual bit is already well established (see the previous sections in this report) and so does not need to be discussed further here. The end tagging and codewords are what enables this protocol to prevent parts of valid message/signature pairs from being used to create forged messages as described in section \ref{sec: ConflictOverIteration}. In the first attack described Bob forged a message by not forwarding the whole message to Charlie thus, changing the meaning of the message. However, as each message bit would have a correct associated private key Charlie would see this as authentic. The end tagging of each encoded message with the bits 11 prevents this. If Charlie receives a signature set which does not begin and end with two signatures representing 1 then he knows it has been altered. As the bits in the message are encoded to 00 or 01 there is no place the message/signature can be "cut" in order to produce the required 11 needed to mark the beginning and end of a valid message. As the bits are numbered and labelled as to what bit they represent the order of them cannot be changed to achieve this either.

\subsection{Further Extensions\label{sec: FurtherExtensions}}
%MDI, multiple parties other QM involved methods. 

Although the quantum digital signature schemes discussed so far are in theory completely secure, many of them have limitations. For example, assumptions made in theoretical models or even laboratory experiments that are required to simplify the problem are detrimental in practical implementations. Assumptions made regarding channel security allow for focus on just malicious actions by just Alice, Bob or Charlie. In reality it will not be the case that only one of those three are attackers. Other considerations include the practical implementation of such schemes and what side channel attacks can be performed on them. These simplifications are necessary in order to create the theoretical models required  but in order for quantum digital signatures to become a viable technology they must be built upon.

\subsubsection{Insecure Channels \label{sec: Insecure Channels}}

With the exception of the QKD based scheme (see section \ref{sec: QKDBasedSignatures}) each of the schemes so far have made the assumption that quantum information is sent over an authenticated quantum channel\cite{Yin16}. Namely that the information sent is always the same as the information received. This simplifies analysis as it ensures that there is no "Eve" (forger intercepting the information). Whilst the same protections against Bob(Charlie) forging would still apply to Eve, there is no way for the recipients to detect that an interception has occurred until the verification step, at which point the individual who received the legitimate signature would disagree with the one that didn't. \\

Whilst there are (costly) methods of implementing an authenticated quantum channel\cite{AdvInQC} this need not always be necessary. First of all we can take a leaf from the book of QKD. As discussed in section \ref{sec: QKDBasedSignatures} Alice and Bob(Charlie) can sacrifice a section of the received QDS to ensure they are sufficiently correlated. If the expected level of error (based on the authentication threshold) is present in this then they can be sure no eavesdropping has occurred. If it is greater than this however, they know Eve is present and can restart the process with a different channel\cite{Amiri16}. \\

The QKD based scheme gives us other methods that can be used in order to bypass the issue of insecure channels for any scheme. In particular that of two separate quantum digital signatures being sent from Bob and Charlie to Alice\cite{Collins16}. Assuming Eve does not intercept both the quantum channels then she does not have access to the whole quantum digital signature. This does not allow for the detection of Eve before the verification step but ensures that so long as she does not have access to at least one quantum channel she cannot commit a forgery. \\

Finally, as proposed with the initial concept for the QKD based QDS scheme is the principle of sending decoy states\cite{Yin16}. A scheme that includes decoy states proceeds with the generation of the two copies of a potential message's quantum signature but before the messages are dispatched they pass through a separate amplitude modulator. This randomly and independently changes the intensity of the signature element pulse to one of three possible values, $\mu$, $\nu$, or $0$. Each with their own defined probability distributions. Only the state in which Bob and Charlie both receive $\mu$ intensity states is deemed the signal state, the other 6 possible combinations are decoy states. Alice then announces the intensity of each pulse allowing Bob and Charlie to discard any states that were decoys.

\subsubsection{Measurement Device Independent \label{sec: MDI}}

As is the case with many concepts in cryptography, a scheme itself may be secure but in its implementation weaknesses may be found and exploited. These are known as side channel attacks. In classical cryptography and example of this would be analysing the power output of a CPU during encryption in order determine further information about the process used. As such in theoretical models and laboratory tests these are often not considered. Side channel attacks are in fact common in the field of quantum cryptography\cite{AdvInQC}\cite{Xu19}. As there is no way to breach the encoding of the information itself, an attacker must look for exploits elsewhere. \\

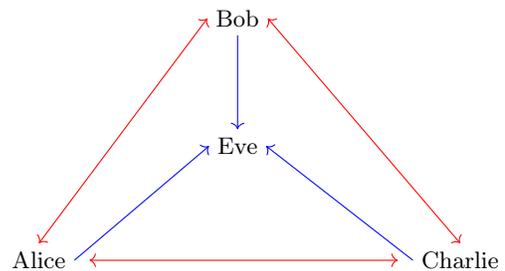
\begin{figure}[h!]
        \begin{center}
        \begin{tikzpicture}[
            ]
           \node[] (a)[] {Alice};
           \node[] (b) [above right = 2.5cm of a, xshift = 0cm, yshift=1cm] {Bob};
           \node[] (c) [right = 4.5cm of a] {Charlie};
           \node[] (e) [below = 1.25cm of b] {Eve};
           
           \draw[blue, ->]   (a.east) -- (e.west);
           \draw[blue, ->]   (c.west) -- (e.east);
           \draw[blue, ->]   (b.south) -- (e.north);
           \draw[red, <->]   (a.north) -- (b.west);
           \draw[red, <->]   ([xshift=0.2cm]a.east) -- ([xshift=-0.2cm]c.west);
           \draw[red, <->]   (b.east) -- (c.north);
           
        \end{tikzpicture}
        \end{center}
        \caption{Representation of the communication channels between Alice, Bob, Charlie and the central relay Eve in MDI QDS schemes. Blue arrows represent quantum communication channels and red represent classical channels. Single headed arrows represent one way communication channels, double headed represent two way channels.}
        \label{fig: MDI}
    \end{figure}

Measurement Device Independent (MDI) schemes bypass these issues by having all quantum communications occur via an untrusted central relay, "Eve"\cite{Puthoor16}, as shown in figure \ref{fig: MDI}. As such none of the other members of the communication actually perform any measurements. They no longer need to be concerned with detector based side channel attacks such as detector blinding as the relay is treated as a "black box". \\

\noindent To do this Alice performs the QKD KGP for QDS technique described in section \ref{sec: QKDBasedSignatures} with Eve. For each state sent, Alice applies random intensity modulation to create a series of decoy states and one signal state (as described in section \ref{sec: Insecure Channels}). This is known as a Measurement Device Independent Key Generation Protocol (MDI KGP) and is based off of MDI QKD protocols\cite{Yin17}\cite{Roberts17}.\\

Eve announces the results of each measurement and their intensity over a public channel. Alice and Bob(Charlie) then communicate over an authenticated classical channel which intensity state was the signal state as well as which basis was used\cite{Puthoor16}. Thus generating a signature without direct quantum communication. Bob and Charlie symmetrise their signature strings and the scheme proceeds to the messaging and verification stages. 

\subsubsection{Expanding to Multiple Parties \label{sec: ExpandingtoMultipleParties}}

\begin{table*}[htbp!]

        \begin{tabular}{c|p{2.5cm}||p{1.5cm}|p{1.6cm}|p{1.5cm}|p{1.5cm}|p{1.5cm}}
            Author & Scheme Summary & Distance (km) & Signature Length & Time to sign (s) & Clock rate (Hz) & Security level \\
            \hline
            \hline
            Collins et al. \cite{Collins14} & Multiport with USE & 0.005 & 5.1$*10^{13}$ & & 100$*10^{6}$ & 10$^{-2}$\\
            \hline
            Collins et al. \cite{Collins16} & DPS QKD & 90 & 2502 & & 10$^{9}$ & 10$^{-4}$ \\
            \hline
            Donaldson et al. \cite{Donaldson16} & USE based post-measurement random forwarding & 0.5 & 1.93$*10^{9}$ & 20 & & 10$^{-2}$ \\
            \hline
            Croal et al. \cite{Croal16} & Hetrodyne Measurements & 1.6 & 7$*10^{4}$ & & 2.2$*10^{6}$ & 10$^{-2}$\\
            \hline
            Yin et al. \cite{Yin172} & Decoy State & 102 & 2.5$*10^{12}$ & 33420 & & 10$^{-5}$ \\
            \hline
            Roberts et al. \cite{Roberts17} & MDI-QKD & 25 & 103336 & 36 & & \\
            \hline
            Yin et al. \cite{Yin17} & MDI-QDS (MDI-KGP) & & 787468 & & &10$^{-7}$\\
            \hline
            Yao et al. \cite{Yao19} & Temporal Ghost Imaging & & 93.9 (signs 10 message bits at a time) & 4 & & 10$^{-4}$

        \end{tabular}
        
        \caption{Summary of the figures of merit of signature schemes referenced in this review. Only includes experimental results and not theoretical estimations of proposed schemes. "Distance" refers to the distance between Alice and Bob or Charlie. "Signature Length" to the bit length of the signature required to sign a single bit message.}
        \label{fig:Key and sig comp table}
    \end{table*}

Each scheme that has been discussed has focused on at maximum three parties communicating. A sender of the message (Alice) and two recipients (Bob and Charlie). In a practical communications scenario this obviously will not be practical\cite{Roberts17}. In the case of sending a mass message authenticated with a quantum digital signature there will be more than two recipients for the same signature. \\

However, this raises two concerns. The first is a practical one; in the schemes described above for a practical quantum communication network, each pair of users will require a quantum communication channel. Scaling as $N(N-1)/2$ links for $N$ users\cite{Roberts17}. As $N$ increases this number becomes less and less practical to implement. A solution to this issue could be the network architecture discussed in section \ref{sec: MDI}. Rather than each pair of members of the communication network having a quantum communication channel between each other they instead each have one quantum channel with an untrusted central relay. This reduces the scaling of the number of required channels to $N-1$ for an $N$ mode network.\\

The second issue is a security concern. If Alice sends out the same quantum digital signature to each recipient then, if there are more than two recipients, for $N$ recipients up to a maximum of $N-1$ of them can collude to attempt to commit forgery against the others. Due to the uncertain nature of the measurements of the quantum signature no single recipient will have a fully correct signature. However, by working together multiple malicious parties can work together to improve their chances of a successful forgery. This issue was first addressed in ref \cite{Arrazola16} where a generic case of a multiparty scheme was first proposed. Ref \cite{Sahin16} further expanded upon this concept. \\

Using the decoy state KGP scheme as a basis (see section \ref{sec: Insecure Channels}) they expand this from only having two recipients to a general case of $N$ recipients\cite{Arrazola16}. To achieve this each recipient generates their own private key for each possible message. They generate a quantum signature for each and distribute to Alice using the process detailed in section \ref{sec: Insecure Channels}. Each recipient then randomly chooses  half of the bits in their private key and forwards it to each other recipient. Resulting in the final private key for each being $(N-1)L/2$. As such from the viewpoint of Alice each private key is exactly the same. Each recipient as well will never know all of the private keys as each other recipient kept half of theirs. Therefore even if $N-1$ recipients colluded they could not successfully forge a signature from Alice. The proposal also discusses the implementation of a system of security levels to quantify how often a signature can be forwarded and remain safe (as this will affect the authentication threshold)\cite{Arrazola16}. On top of this is a majority voting protocol for resolving disputes. Thus, fully outlining the protocols required to generalise a QDS protocol to any number of recipients.

\subsection{Concluding Remarks on QDS\label{sec: Concluding Remarks}}
    
The field of quantum digital signatures is a relatively new area of research but in the time it has existed it has developed from a purely theoretical concept into (albeit limited) practical implementation over 100km of fibre\cite{Yin172}. Arguably the most important advance that allowed for feats such as this is that of random forwarding. Whilst simplistic in design this has allowed for the reduction of the quantum technology required in QDS down to simply that used in the already well tested QKD. This paved the way for future schemes which improved upon the basic random forwarding scheme detailed in section \ref{sec: RandomForwarding}. This was achieved either with better hardware or by using it as a primitive for new concepts such as the QKD KGP scheme, which at the moment boasts one of the quickest signing rates over the longest distance.\\

Quantum digital signatures however, are by no means ready for full commercial use. No current scheme fully solves all of the issues that would allow for QDS to move into broader practical use. For a scheme to be viable for use it must not rely on the assumptions stated in many papers to work. It must be able to work with insecure channels, be immune to side channel attacks and allow for both multiple bits and multiple users securely. The MDI KGP scheme detailed in section \ref{sec: ExpandingtoMultipleParties} (with the recipients sending the signatures to Alice) currently comes the closest to achieving this. It is secure against forging and repudiation and if adapted to include the multi-bit signing techniques detailed in section \ref{sec: MultiBit Signing} would allow for many bit messages to be sent to many recipients. No one has yet, however, attempted to practically implement such a scheme.

\section{Conclusion}
\noindent Signatures play a vital role in the security and trustworthiness of communications and, moving forwards, there are valid concerns about the long-term reliability of the digital signature schemes that have been widely adopted. Solutions basing their security on quantum mechanics, rather than complex mathematics are appealing but are far from commercial readiness, and post-quantum digital signatures form an optimal middle ground whilst quantum technologies mature.

\bibliographystyle{unsrt}
\bibliography{main.bib}

\end{document}